\documentclass[fleqn,usenatbib]{mnras}

\usepackage[T1]{fontenc}
\usepackage{ae,aecompl}

\usepackage[figurename=Figure.,
                  justification=RaggedRight, 
                  labelfont={bf, footnotesize}, 
                  textfont={footnotesize},position=top]{caption}

\usepackage{graphicx}	
\usepackage{amsmath}	
\usepackage{amssymb}	
\usepackage{float}
\usepackage{dblfloatfix}
\usepackage{blindtext}
\usepackage{booktabs}
\usepackage{rotating}
\usepackage{natbib}
\usepackage{fmtcount}
\usepackage{threeparttablex}
\usepackage{multirow}
\usepackage{morefloats}
\usepackage{dblfloatfix}
\usepackage{dcolumn}
\usepackage{amsmath}
\usepackage{latexsym}
\usepackage{longtable}
\usepackage{lipsum} 
\usepackage{textcomp}
\usepackage{gensymb}
\usepackage{mathtools}
\usepackage{afterpage}
\usepackage{caption}

\title[]{SN 2015an: a normal luminosity type II supernova with low expansion velocity at early phases}

\author[Raya Dastidar et al.]{Raya Dastidar$^{1,2}$\thanks{E-mail: rayadastidar@aries.res.in, rdastidr@gmail.com},
Kuntal Misra$^{1,3}$, Stefano Valenti$^{3}$, Jamison Burke$^{4,5}$, \newauthor Griffin Hosseinzadeh$^{6}$, Anjasha Gangopadhyay$^{1,7}$, D. Andrew Howell$^{4,5}$, \newauthor Mridweeka Singh$^{1,7}$, Iair Arcavi$^{8}$, Brijesh Kumar$^{1}$, Curtis McCully$^{4,5}$, \newauthor Pankaj Sanwal$^{1,7}$ and S. B. Pandey$^{1}$
\\
$^{1}$Aryabhatta Research Institute of observational sciencES, Manora Peak, Nainital 263 001 India\\
$^{2}$Department of Physics \& Astrophysics, University of Delhi, Delhi-110 007\\
$^{3}$Department of Physics, University of California, 1 Shields Ave, Davis, CA 95616-5270, USA\\
$^{4}$Las Cumbres Observatory, 6740 Cortona Dr., Suite 102, Goleta, CA 93117-5575, USA\\
$^{5}$Department of Physics, University of California, Santa Barbara, CA 93106-9530, USA\\
$^{6}$Center for Astrophysics | Harvard \& Smithsonian, 60 Garden Street, Cambridge, MA 02138-1516, USA\\
$^{7}$Pt.Ravi Shankar Shukla University, Raipur 492 010,  India\\
$^{8}$The School of Physics and Astronomy, Tel Aviv University, Tel Aviv 69978, Israel
}
\date{Accepted XXX. Received YYY; in original form ZZZ}

\pubyear{2018}

\begin{document}

\label{firstpage}
\pagerange{\pageref{firstpage}--\pageref{lastpage}}
\maketitle

\begin{abstract}
We present the photometry and spectroscopy of SN~2015an, a Type II Supernova (SN) in IC~2367. The recombination phase of the SN lasts up to $\sim$120~d, with a decline rate of 1.24~mag/100d, higher than the typical SNe IIP. The SN exhibits bluer colours than most SNe II, indicating higher ejecta temperatures. The absolute $V$-band magnitude of SN~2015an at 50~d is $-$16.83$\pm$0.04~mag, pretty typical for SNe II. However, the $^{56}$Ni mass yield, estimated from the tail $V$-band light curve to be 0.021$\pm$0.010~M$_\odot$, is comparatively low. The spectral properties of SN~2015an are atypical, with low H$\alpha$ expansion velocity and presence of high velocity component of H$\alpha$ at early phases. Moreover, the continuum exhibits excess blue flux up to $\sim$50~d, which is interpreted as a progenitor metallicity effect. The high velocity feature indicates ejecta-circumstellar material interaction at early phases. The semi-analytical modelling of the bolometric light curve yields a total ejected mass of $\sim$12~M$_\odot$, a pre-supernova radius of $\sim$388~R$_\odot$ and explosion energy of $\sim$1.8 foe.
\end{abstract}




\section{Introduction}
\label{sec1}
The hydrogen-dominant class of supernovae (SNe) designated as SNe~II, are the fate of the majority of massive stars ($\gtrsim$~8~M$_\odot$) which ensue from the gravitational collapse of the iron core in the red supergiant (RSG) stage \citep{2009MNRAS.395.1409S, 2015PASA...32...16S}. SNe II are characterized by the presence of Balmer lines in their early spectra \citep{1941PASP...53..224M}, and those with abundant hydrogen either exhibit a plateau (P) or a linearly declining phase (L) in their light curve which lasts for about 100 days, before plummeting into the radioactive tail phase \citep{1979A&A....72..287B}. Three other subtypes exist: Type IIn, IIb and the 1987A-like, but hereafter we will mainly discuss SNe IIP and IIL conjointly as SNe~II. Direct evidence for RSG as progenitor of SNe~II are proclaimed from fortuitous detection of the progenitor in pre-explosion images, which corroborated with the theoretical predictions of \cite{1971Ap&SS..10...28G, 1976ApJ...207..872C, 1977ApJS...33..515F}. The initial mass range, however, is constrained between 7-18~M$_\odot$ from the pre-explosion images, that raised the \textquoteleft RSG problem\textquoteright{} \citep{2009MNRAS.395.1409S, 2015PASA...32...16S}, as RSGs more massive than 18~M$_\odot$ are known to exist observationally. This is further substantiated by the higher progenitor mass yields from hydrodynamical modelling of the SN observables (e.g. \citealt{2017MNRAS.464.3013P,2018MNRAS.479.2421D} and references therein). But ejecta masses of SNe~II estimated from hydrodynamical modelling of light curve may not be robust and unique, as indicated in recent simulation of SNe~II explosions by \cite{2019A&A...625A...9D}.

The remarkable feature of the SNe II population is the continuity in diversity. Despite the prominent dispersion in the observed properties of SNe~II, which is apparent from the range of peak magnitudes ($-$14 $\gtrsim$ M$_B$ $\gtrsim$ $-$18; \citealt{1994A&A...282..731P}), plateau luminosities ($-$15 \textgreater~M$_V$ \textgreater~$-$18; \citealt{2003ApJ...582..905H}) and ejecta velocities (1500 \textless~v$_{ph}^{50}$ \textless~9600 km s$^{-1}$; \citealt{2017ApJ...850...89G}), a continuum in the properties has been noted \citep{2014ApJ...786...67A,2015ApJ...799..208S}. The progenitor and explosion properties, such as the radius of the progenitor, the mass of hydrogen envelope and the synthesized $^{56}$Ni mass, most likely gives rise to the continuity. External factors, such as the  presence of a dense circumstellar (CS) shell in proximity to the progenitor star at the time of explosion could also be responsible for the observed continuum in the properties of SNe II \citep{2017ApJ...838...28M}.

The spectra of SNe II exhibits broad P~Cygni lines, which seems to disfavour the possibility of a major contribution of circumstellar interaction (CSI) of ejecta in powering these explosions, unlike SNe IIn, which are characterised by narrow emission lines of hydrogen. However, the interaction in SNe~II can be hidden below the photosphere, and not be seen in narrow lines (e.g. iPTF14hls, \citealt{2018MNRAS.477...74A}), and still contribute to the luminosity. A contribution of interaction to the luminosity, with spectra bereft of narrow lines, is also possible for certain shallow wind density profiles \citep{2012ApJ...747..118M}. The spectra of low-luminosity ($-$14~\textgreater~M$_V^{max}~$\textgreater~$-$15.5) class of SNe II (e.g. SN~2005cs) exhibit weak absorption features superimposed on a dominant blue continuum lasting up to $\lesssim$ 30~d post explosion \citep{2014MNRAS.439.2873S}, owing to the underenergetic nature of the explosion. However, normal luminosity SNe showing weak absorption component (e.g. LSQ13fn, \citealt{2016A&A...588A...1P}) are most likely the terminal explosion of a sub-solar metallicity progenitor \citep{2013MNRAS.433.1745D}. 

Despite the surge in the number of studies undertaken for individual as well as samples of SNe II, a number of issues remain inconclusive, such as the source of the observed diversity in their photometric and spectroscopic properties. Dust formation at late times and CSM forged in the latest stages of the evolution can considerably tweak the observed properties of the explosion. To enrich our understanding of the inhomogeneous class of SNe II, detailed studies of these events particularly the deviant objects are important. 

In this paper, we present the photometric and spectroscopic analysis of a Type II SN, SN~2015an, discovered by Berto Monard (Bronberg Observatory) in the galaxy IC~2367 on 2015 September 13.15~UT (JD 2457278.65) at an unfiltered magnitude of 15.2~mag. The classification spectrum was procured by the Las Cumbres Observatory (LCO) Supernova Key Project on 2015 September~26.7~UT \citep{2015ATel.8102....1H}, nearly two weeks from discovery, with the robotic FLOYDS instrument mounted on the Faulkes Telescope South (FTS). The spectrum exhibited a striking blue continuum and a significantly low H$\alpha$ expansion velocity ($\sim$ 5000~km~s$^{-1}$). The predominance of blue continuum in spectra up to two weeks past discovery is generally observed in low-luminosity SNe, however, the expansion velocity of H$\alpha$ and the luminosity of SN~2015an are comparatively higher than the low-luminosity events (see Figs.~\ref{abs_lc} \& \ref{fig:vel_comp}). This led to its classification as a peculiar SN II. The details of SN~2015an and its host galaxy IC~2367 are given in Table~\ref{tab:sn15an_IC2367_detail}. 

\begin{table}
\centering
 \begin{minipage}{84mm}
\caption{Basic information on SN 2015an and the host galaxy IC~2367. The host galaxy parameters are taken from NED.}
\begin{tabular}{@{}cc@{}}
\hline
Host galaxy & IC~2367   \\
Galaxy type & SBb \\  
Redshift & 0.00817 $\pm$ 0.00001$^\dagger$ \\ 
Major Diameter & 2.4 arcmin \\
Minor Diameter & 1.7 arcmin \\
Helio. Radial Velocity &  2488 $\pm$ 3 km s$^{-1}$ \\
\hline
Offset from nucleus & 71$^{\prime\prime}$.0 E,4$^{\prime\prime}$.0 N \\
Distance & 30.5 $\pm$ 0.6 Mpc$^{\dagger\dagger}$ \\
Total Extinction E(B-V) & 0.0875 $\pm$ 0.0012 mag$^{\dagger\dagger}$ \\
SN type & II\\
Date of Discovery & 2457278.65 (JD) \\
Estimated date of explosion & 2457268.5$\pm$1.6 (JD) $^{\dagger\dagger}$\\
\hline 
\end{tabular}
\newline
$^\dagger$ \citet{1998A&AS..130..333T}
$^{\dagger\dagger}$ This paper.       
\label{tab:sn15an_IC2367_detail}   
   \end{minipage}
\end{table}

The Virgo infall distance to the galaxy IC~2367 is 30.7~$\pm$~0.7~Mpc, where we have used the recessional velocity of the galaxy v$_{Vir}$~=~2259~$\pm$~3~km~s$^{-1}$ from HyperLeda \citep{2014AA...570A..13M} and Hubble constant H$_0$~=~73.48~$\pm$~1.66~km~s$^{-1}$~Mpc$^{-1}$ \citep{2018ApJ...855..136R}. We used the expanding photosphere method (EPM) to estimate the distance to SN~2015an (details in Sect. \ref{dist}), which yielded 29.8~$\pm$~1.5~Mpc. The weighted mean of the two estimations is 30.5$\pm$0.6~Mpc, which we have adopted as distance to SN~2015an in this paper. EPM also estimates the explosion epoch to be 2457268.5 $\pm$ 1.6 d (2015~September~03~UT) which we will refer to as 0~d.

The Galactic reddening E(B$-$V)$_{MW}$ in the line of sight of IC~2367 is 0.0875 $\pm$ 0.0012 mag, obtained from the extinction dust maps of \cite{2011ApJ...737..103S}. In our best resolution optical spectra, we do not detect the absorption features due to interstellar Na~{\sc i}~D lines from the Galaxy. From a set of best SNR spectra, we constructed a composite spectrum following the prescription in \cite{2015A&A...582A...3G} to determine the limiting equivalent width (EW) of Na~{\sc i}~D absorption feature at the redshift of the host galaxy. Gaussian profile of varying EWs were fitted to the Na~{\sc i}~D absorption line in the stacked spectrum and plausibly a weak Na~{\sc i}~D line with EW of 0.2~\AA{} could be discerned. This corresponds to a host galaxy reddening of \textless 0.02 mag (using eqn (9) of \citealt{2012MNRAS.426.1465P}). The weak Na~{\sc i}~D feature indicates negligible reddening contribution from the host, which is in agreement with the remote location of the SN from the main body of its host galaxy. Consequently, we infer that the total reddening is arising only from the Galactic component, that is E(B $-$ V)$_{tot}$ = 0.0875 $\pm$ 0.0012 mag and is used in the rest of our analysis.

The paper is structured as follows. Sect.~\ref{sec2} gives details of the data and reduction procedure. We investigate the photometric and spectroscopic properties of SN~2015an in Sect.~\ref{sec4} and Sect.~\ref{sec5}, respectively. The distance derived using the expanding photosphere method is elaborated in Sect.~\ref{dist}. The modelling of the bolometric light curve using semi-analytic methods is discussed in Sect.~\ref{sec7}. Finally, we examine the overall properties of SN~2015an in Sect.~\ref{sec8} and present a short summary of the paper in Sect.~\ref{sec9}.

\section{SN 2015an: Photometry and Spectroscopy}
\label{sec2}

\begin{figure}
\begin{center}
\includegraphics[scale=0.4,clip, trim={0cm 0cm 0cm 0cm}]{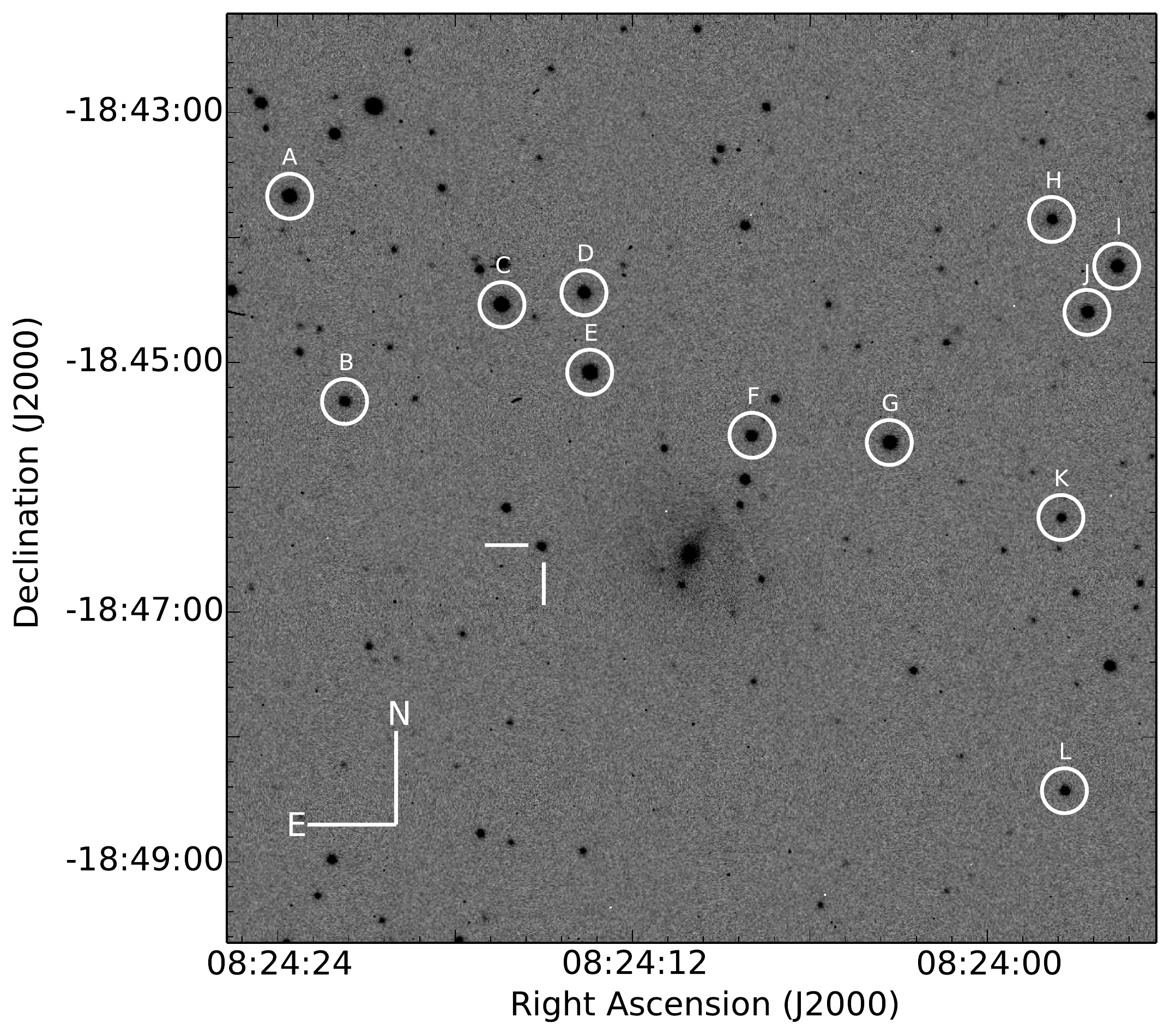}
\end{center}
\caption{SN 2015an in IC 2367, star ID for local standards have been marked.}
\label{fig:local}
\end{figure}

\begin{table*}
\caption{Summary of the instruments used for the observational campaign of SN 2015an.}
\centering
\smallskip
\flushleft

\begin{tabular}{l l l l l l l}
\hline

Telescope & Location & Instrument  & Pixel Scale    & Imaging & Dispersers/ Grisms   &   \\
                        &  &   & ($''$/pixel) & bands\\
\hline
1.04m Sampurnanand & ARIES Observatory, India & Tek 1k$\times$1k   & 0.53  & {\it VRI}  & - & 1\\
Telescope (ST)         &  &  & & \\
1.3m Devasthal Fast & ARIES Observatory, India & Andor 512$\times$512 & 0.64  & {\it BVRI} & - & 2\\
Optical Telescope (DFOT)  &  &  & & \\
1m LCO & Siding Spring Observatory & Sinistro & 0.389 & $BVg^\prime r^\prime i^\prime$ & - & 3 \\
1m LCO  & South African  & Sinistro & 0.389    & $BVg^\prime r^\prime i^\prime$ & -  & 4\\
& Astronomical Observatory & & & \\
1m LCO & Cerro Tololo & Sinistro  & 0.389 &  $BVg^\prime r^\prime i^\prime$   & - & 5\\
& Interamerican Observatory & & & \\
1m LCO & McDonald Observatory, USA & Sinistro & 0.389 & $BVg^\prime r^\prime i^\prime$ & - & 6\\
2m FTN & Haleakala Observatory, USA & Spectral, FLOYDS & 0.304, 0.34 & $g^\prime r^\prime i^\prime$ & Cross disperser & 7\\
2m FTS & Siding Spring Observatory, & FLOYDS & 0.34 & - & Cross disperser & 8\\
& Australia & & & \\
\hline                                   
\end{tabular}

\label{tab:details_instrument_detectors}      
\end{table*}

\begin{table*}
\caption{Coordinates and photometry of the local sequence reference stars in the $BVg^{\prime}r^{\prime}i^{\prime}$ bands.}
\begin{tabular}{llcccccc}
\hline
 ID & $ \alpha_{J2000.0} $ & $ \delta_{J2000.0} $ & $B$ & $V$  & $g^\prime$  & $r^\prime$ &$i^\prime$ \\
 & (hh:mm:ss) &(dd:mm:ss) & mag & mag & mag & mag & mag \\
\hline 
A & 08:24:23.6 & $-$18:43:40.2 & 14.227 (.015) & 13.729 (.021) & 13.928 (.032) & 13.594 (.019) & 13.503 (.014) \\
B & 08:24:21.7 & $-$18:44:26.6 & 15.428 (.028) & 14.866 (.037) & 15.095 (.027) & 14.717 (.024) & 14.580 (.033) \\
C & 08:24:16.4 & $-$18:45:18.8 & 14.752 (.029) & 13.458 (.024) & 14.056 (.033) & 12.999 (.023) & 12.574 (.015) \\
D & 08:24:13.6 & $-$18:43:54.4 & 14.926 (.027) & 14.334 (.009) & 14.588 (.035) & 14.184 (.016) & 14.017 (.046) \\
E & 08:24:13.4 & $-$18:45:04.7 & 14.220 (.025) & 13.553 (.018) & 13.827 (.034) & 13.347 (.024) & 13.174 (.018) \\
F & 08:24:07.9 & $-$18:45:35.1 & 15.489 (.040) & 14.613 (.025) & 14.996 (.036) & 14.295 (.019) & 14.035 (.047) \\
G & 08:24:03.3 & $-$18:47:24.2 & 14.724 (.022) & 13.749 (.015) & 14.194 (.034) & 13.446 (.015) & 13.129 (.026) \\
H & 08:23:57.8 & $-$18:43:51.3 & 15.516 (.052) & 15.071 (.026) & 15.233 (.038) & 14.936 (.010) & 14.927 (.038) \\
I & 08:23:55.6 & $-$18:44:13.7  & 14.946 (.034) & 14.086 (.018) & 14.474 (.035) & 13.832 (.016) & 13.592 (.016) \\
J & 08:23:56.6 & $-$18:44:35.9 & 14.492 (.033) & 14.330 (.022) & 14.332 (.039) & 14.381 (.028) & 14.543 (.028) \\
K & 08:23:57.5 & $-$18:46:14.5 & 16.709 (.054) & 15.575 (.031) & 16.122 (.038) & 15.089 (.041) & 14.759 (.047) \\
L & 08:27:57.4 & $-$18:48:25.8 & 16.138 (.031) & 15.31 (.016) & 15.680 (.042) & 15.039 (.030) & 14.840 (.044) \\

\hline

\end{tabular}
\label{tab:local}   
\end{table*}

The photometric campaign of SN~2015an was triggered on the day of its discovery using instruments equipped with broadband $BVRI$ and $g^\prime$$r^\prime$$i^\prime$$z^\prime$ filters as listed in Table \ref{tab:details_instrument_detectors}. The photometric observations continued up to 261~d from explosion. The pre-processing of the images including bias and flat-field corrections were carried out in IRAF\footnote{IRAF stands for Image Reduction and Analysis Facility distributed by the National Optical Astronomy Observatories which is operated by the Association of Universities for research in Astronomy, Inc., under cooperative agreement with the National Science Foundation.} environment and cosmic rays were cleaned using the {\sc l.a.cosmic} routine \citep{2001PASP..113.1420V}. Owing to the remoteness of SN 2015an from its host galaxy nucleus (at a deprojected radial distance of 13.9~kpc), we performed the point spread function photometry using {\sc daophot~ii} \citep{1987PASP...99..191S} to derive the instrumental magnitudes. The LCO photometry was carried using \texttt{lcogtsnpipe} designed by Stefano Valenti (see details in \citealt{2016MNRAS.459.3939V}). The instrumental magnitudes were calibrated using the standard magnitudes of 12 local standard stars in the SN field obtained from the AAVSO Photometric All-Sky Survey (APASS\footnote{https://www.aavso.org/apass}) catalog. The local standards are marked in Fig. \ref{fig:local} and their magnitudes are listed in Table \ref{tab:local}. The calibrated SN magnitudes, thus obtained, are listed in Table~\ref{photometry}. 

\begin{table*}
 \caption{Optical photometry of SN 2015an.}
\centering
\begin{tabular}{c c c c c c c c c}
\hline
UT Date    & JD & Phase$^\dagger$ & {\textit{B}} &  {\textit{V}}        & {\textit{g$^\prime$}}         & {\textit{r$^\prime$}}         &  {\textit{i$^\prime$}}        & {\textit{Tel}} \\
(yyyy-mm-dd) & 2457000+ & (days)  & (mag)        &   (mag)              & (mag)                & (mag)                &   (mag)              & \\
\hline
2015-09-13.8 & 279.3 &  10.8  &  -                   &  -                   &  -                   & 15.56 $\pm$ 0.02 & 15.75 $\pm$ 0.01 & 3 \\
2015-09-14.1 & 279.6 &  11.1 &  -                   &  -                   & 15.37 $\pm$ 0.01 & 15.63 $\pm$ 0.02 & 15.78 $\pm$ 0.02 & 4 \\
2015-09-15.8 & 281.3 &  12.8 & 15.59 $\pm$ 0.01 & 15.52 $\pm$ 0.01 & -                    & -                    & -                    & 3 \\
2015-09-16.1 & 281.6 &  13.1 & -                    & -                    & 15.48 $\pm$ 0.01 & 15.63 $\pm$ 0.01 & 15.82 $\pm$ 0.02 & 4 \\
2015-09-18.4 & 283.9 &  15.4 & -                    & -                    & 15.62 $\pm$ 0.01 & 15.65 $\pm$ 0.01 & 15.74 $\pm$ 0.01 & 5 \\
2015-09-19.4 & 284.9 &  16.4 &  -                   & -                    & 15.56 $\pm$ 0.08 & 15.60 $\pm$ 0.01 & 15.72 $\pm$ 0.01 & 5 \\
2015-09-20.8 & 286.2 &  17.7 &  -                   &  -                   & 15.48 $\pm$ 0.10 & 15.58 $\pm$ 0.01 & 15.67 $\pm$ 0.01 & 3 \\
2015-09-21.1 & 286.6 &  18.1 &  -                   & -                    & 15.40 $\pm$ 0.01 & 15.54 $\pm$ 0.01 & 15.67 $\pm$ 0.01 & 4 \\
2015-09-21.4 & 286.9 &  18.4 & 15.60 $\pm$ 0.01 & 15.47 $\pm$ 0.01 &  -                   & -                    & -                    & 5 \\
2015-09-22.4 & 287.9 &  19.4 & 15.64 $\pm$ 0.01 & 15.49 $\pm$ 0.01 &  -                   & 15.54 $\pm$ 0.01 & 15.62 $\pm$ 0.01 & 5 \\
2015-09-26.4 & 291.8 &  23.3 & 15.75 $\pm$ 0.02 & 15.56 $\pm$ 0.02 & 15.55 $\pm$ 0.01 & 15.57 $\pm$ 0.01 & 15.65 $\pm$ 0.02 & 5 \\
2015-10-01.8 & 297.3 &  28.8 & 15.89 $\pm$ 0.01 & 15.58 $\pm$ 0.01 & 15.66 $\pm$ 0.01 & 15.52 $\pm$ 0.01 & 15.52 $\pm$ 0.01 & 3 \\
2015-10-05.7 & 301.2 &  32.7 & 16.03 $\pm$ 0.01 & 15.66 $\pm$ 0.01 & 15.75 $\pm$ 0.01 & 15.57 $\pm$ 0.01 & 15.59 $\pm$ 0.01 & 3 \\
2015-10-13.7 & 309.2 &  40.7 & 16.34 $\pm$ 0.02 & 15.77 $\pm$ 0.01 & 16.00 $\pm$ 0.01 & 15.64 $\pm$ 0.01 & 15.60 $\pm$ 0.02 & 5 \\
2015-10-17.3 & 312.8 &  44.3 & 16.52 $\pm$ 0.02 & 15.87 $\pm$ 0.01 & 16.14 $\pm$ 0.01 & 15.75 $\pm$ 0.01 & 15.67 $\pm$ 0.02 & 6 \\
2015-10-26.5 & 321.9 &  53.4 & 16.87 $\pm$ 0.02 & 16.06 $\pm$ 0.02 & -                    & 15.86 $\pm$ 0.01 & -                    & 4 \\
2015-10-31.1 & 326.6 &  58.1 & 16.95 $\pm$ 0.03 & 16.12 $\pm$ 0.02 & 16.47 $\pm$ 0.02 & 15.92 $\pm$ 0.01 & 15.81 $\pm$ 0.02 & 4 \\
2015-11-05.1 & 331.6 &  63.1 & 17.15 $\pm$ 0.04 & 16.18 $\pm$ 0.02 & 16.58 $\pm$ 0.02 & 15.99 $\pm$ 0.04 & -                    & 4 \\
2015-11-09.1 & 335.6 &  67.1 & -                    & 16.24 $\pm$ 0.02 & 16.63 $\pm$ 0.01 & 16.03 $\pm$ 0.01 & 15.92 $\pm$ 0.02 & 4 \\
2015-11-09.9 & 336.4 & 67.9 & -                    & 16.38 $\pm$ 0.01 & - & 15.94 $\pm$ 0.01 & 15.96 $\pm$ 0.01 & 1\\
2015-11-10.9 & 337.4 & 68.9 & -                    & 16.35 $\pm$ 0.01 & - & 16.02 $\pm$ 0.01 & 15.91 $\pm$ 0.01 & 1\\
2015-11-13.3 & 339.8 & 71.3 & -                    & 16.28 $\pm$ 0.02 & -                    & 16.02 $\pm$ 0.02 & 16.01 $\pm$ 0.04 & 5 \\
2015-11-16.9 & 343.4 & 74.9 & 17.38 $\pm$ 0.04 & 16.28 $\pm$ 0.02 & 16.73 $\pm$ 0.02 & 16.03 $\pm$ 0.01 & 15.95 $\pm$ 0.02 & 4 \\
2015-11-20.7 & 347.2 & 78.7 & 17.32 $\pm$ 0.04 & 16.31 $\pm$ 0.02 & 16.78 $\pm$ 0.02 & 16.04 $\pm$ 0.01 & 15.93 $\pm$ 0.02 & 3 \\
2015-11-22.6 & 349.1 & 80.6 & 17.46 $\pm$ 0.04 & 16.34 $\pm$ 0.02 & 16.91 $\pm$ 0.02 & 16.10 $\pm$ 0.02 & 16.00 $\pm$ 0.02 & 3 \\
2015-11-24.1 & 350.6 & 82.1 & 17.47 $\pm$ 0.04 & 16.41 $\pm$ 0.02 & 16.89 $\pm$ 0.02 & 16.12 $\pm$ 0.01 & 16.04 $\pm$ 0.02 & 4 \\
2015-11-27.9 & 354.4 & 85.9 & 17.39 $\pm$ 0.05 & 16.49 $\pm$ 0.03 & 16.94 $\pm$ 0.02 & 16.18 $\pm$ 0.02 & 16.06 $\pm$ 0.02 & 4 \\
2015-12-05.9 & 362.4 & 93.9 & 17.64 $\pm$ 0.03 & 16.60 $\pm$ 0.01 & & 16.28 $\pm$ 0.01 & 16.15 $\pm$ 0.01 & 2\\
2015-12-06.9 & 363.4 & 94.9 & -                    & 16.64 $\pm$ 0.03 & 17.07 $\pm$ 0.02 & 16.24 $\pm$ 0.02 & 16.15 $\pm$ 0.03 & 4 \\
2015-12-10.7 & 367.1 & 98.6 & 17.74 $\pm$ 0.03 & 16.59 $\pm$ 0.02 & 17.15 $\pm$ 0.02 & 16.34 $\pm$ 0.01 & 16.21 $\pm$ 0.02 & 3 \\
2015-12-14.9 & 371.5 & 103.0 & 17.87 $\pm$ 0.05 & 16.74 $\pm$ 0.03 & 17.28 $\pm$ 0.02 & 16.42 $\pm$ 0.02 & 16.32 $\pm$ 0.02 & 4 \\
2015-12-18.4 & 374.9 & 106.4 & 17.99 $\pm$ 0.04 & 16.84 $\pm$ 0.02 & 17.32 $\pm$ 0.02 & 16.46 $\pm$ 0.02 & 16.36 $\pm$ 0.02 & 6 \\
2015-12-22.1 & 378.6 & 110.1 & 18.20 $\pm$ 0.07 & 16.96 $\pm$ 0.02 & -                    & 16.55 $\pm$ 0.02 & 16.44 $\pm$ 0.04 & 5 \\
2015-12-23.9 & 380.4 & 111.9 & - & 17.22 $\pm$ 0.02 & - & 16.66 $\pm$ 0.01 & 16.57 $\pm$ 0.01 & 1\\
2015-12-25.8 & 382.4 & 113.9 & -                    & -                    & -                    & -                    & 16.54 $\pm$ 0.04 & 4 \\
2015-12-29.6 & 386.1 & 117.6 & 18.43 $\pm$ 0.15 & -                    & 17.56 $\pm$ 0.03 & -                    & -                    & 3 \\
2015-12-30.7 & 387.2 & 118.7 & - & 17.18 $\pm$ 0.02 & - & 16.68 $\pm$ 0.01 & 16.62 $\pm$ 0.01 & 1\\
2016-01-01.8 & 389.3 & 120.8 & - & 17.31 $\pm$ 0.02 & - & 16.90 $\pm$ 0.01 & 16.76 $\pm$ 0.01 & 1\\
2016-01-02.7 & 390.2 & 121.7 & - & 17.27 $\pm$ 0.01 & - & 16.80 $\pm$ 0.01 & 16.75 $\pm$ 0.01 & 1\\
2016-01-02.9 & 390.4 & 121.9 & 18.70 $\pm$ 0.10 & 17.26 $\pm$ 0.05 & -                    & -                    & 16.64 $\pm$ 0.03 & 4 \\
2016-01-03.8 & 391.3 & 122.8 & - & - & - & 16.85 $\pm$ 0.01 & 16.79 $\pm$ 0.01 & 1\\
2016-01-05.7 & 393.2 & 124.7 & -              & 17.42 $\pm$ 0.01 & - & 16.92 $\pm$ 0.01 & 16.87 $\pm$ 0.01 & 1\\
2016-01-08.1 & 395.6 & 127.1 & 18.93 $\pm$ 0.10 & 17.65 $\pm$ 0.04 & 18.38 $\pm$ 0.04 & 17.10 $\pm$ 0.03 & 17.07 $\pm$ 0.04 & 5 \\
2016-01-11.5 & 399.0 & 130.5 & 19.56 $\pm$ 0.12 & 18.31 $\pm$ 0.06 & 19.06 $\pm$ 0.10 & 17.74 $\pm$ 0.04 & 17.64 $\pm$ 0.06 & 3 \\
2016-01-12.7 & 400.2 & 131.7 & 19.94 $\pm$ 0.04 & 18.64 $\pm$ 0.01 & - & 18.16 $\pm$ 0.02 & 17.97 $\pm$ 0.02 & 2\\
2016-01-13.7 & 401.2 & 132.7 & 20.17 $\pm$ 0.06 & 19.01 $\pm$ 0.02 & - & 18.16 $\pm$ 0.01 & 18.14 $\pm$ 0.01 & 2\\
2016-01-15.8 & 403.3 & 134.8 & 20.41 $\pm$ 0.06 & 19.34 $\pm$ 0.02 & - & 18.50 $\pm$ 0.01 & 18.13 $\pm$ 0.01 & 2\\
2016-01-16.8 & 404.2 & 135.7 & 20.49 $\pm$ 0.12 & 19.40 $\pm$ 0.06 & - & 18.53 $\pm$ 0.02 & 18.31 $\pm$ 0.02 & 2\\
2016-01-20.2 & 407.7 & 139.2 &  -                   & 19.68 $\pm$ 0.01 & 20.13 $\pm$ 0.09 &       -              & 18.56 $\pm$ 0.10 & 5 \\
2016-01-30.6 & 418.1 & 149.6 & - & 19.71 $\pm$ 0.05 & - & 18.98 $\pm$ 0.02 & 18.79 $\pm$ 0.03 & 2\\
2016-01-31.7 & 419.2 & 150.7 & 20.97 $\pm$ 0.16 & 19.60 $\pm$ 0.04 & - & 18.98 $\pm$ 0.02 & 18.90 $\pm$ 0.02 & 2\\
2016-02-03.8 & 422.3 & 153.8 & 20.86 $\pm$ 0.28 & 19.90 $\pm$ 0.08 & 20.34 $\pm$ 0.09 & 18.86 $\pm$ 0.03 & 18.76 $\pm$ 0.09 & 4 \\
2016-02-08.2 & 426.7 & 158.2 & 21.12 $\pm$ 0.15 & 19.80 $\pm$ 0.05 & 20.44 $\pm$ 0.10 & 19.03 $\pm$ 0.05 & 19.00 $\pm$ 0.06 & 5 \\
2016-02-10.7 & 429.2 & 160.7 & - & - & - & 18.92 $\pm$ 0.03 & 18.94 $\pm$ 0.04& 1\\
2016-02-10.7 & 429.2 & 160.7 & - & 19.68 $\pm$ 0.03 & - & 18.80 $\pm$ 0.02 & 18.66 $\pm$ 0.02 & 2\\
2016-02-11.7 & 430.2 & 161.7 & - & 19.70 $\pm$ 0.06 & - & 18.94 $\pm$ 0.02 & 18.90 $\pm$ 0.04 & 2\\
2016-02-11.7 & 430.2 & 161.7 & - & - & - & 18.90 $\pm$ 0.03 & 18.97 $\pm$ 0.04 & 1\\
2016-02-12.7 & 431.2 & 162.7 & 21.12 $\pm$ 0.15 & 19.88 $\pm$ 0.05 & - & 19.00 $\pm$ 0.02 & 18.87 $\pm$ 0.03 & 2\\
2016-02-14.2 & 432.7 & 164.2 & 20.95 $\pm$ 0.02 & 19.86 $\pm$ 0.11 & 20.45 $\pm$ 0.15 & 19.14 $\pm$ 0.03 & 19.19 $\pm$ 0.05 & 5 \\ 
2016-02-26.6 & 445.1 & 176.6 & - & - & - & 19.07 $\pm$ 0.02 & 19.04 $\pm$ 0.03 & 1\\
2016-02-27.7 & 446.2 & 177.7 & - & - & - & 19.10 $\pm$ 0.03 & 19.00 $\pm$ 0.02 & 1\\
2016-03-01.7 & 449.2 & 180.7 & - & - & - & 18.93 $\pm$ 0.02 & 18.99 $\pm$ 0.03 & 1\\
2016-03-02.7 & 450.2 & 181.7 & - & - & - & 19.02 $\pm$ 0.02 & 19.02 $\pm$ 0.03 & 1\\
2016-03-05.4 & 452.9 & 184.4 & -                    & -                    & 20.49 $\pm$ 0.03 & 18.99 $\pm$ 0.01 & 19.28 $\pm$ 0.03 & 7 \\ 
2016-04-10.3 & 488.7 & 220.2 & -                    & -                    & 20.76 $\pm$ 0.04 & -                    & 19.71 $\pm$ 0.02 & 7 \\ 
2016-05-20.9 & 529.5 & 261.0 & -                    & -                    & 20.98 $\pm$ 0.19 & 19.92 $\pm$ 0.07 & 20.76 $\pm$ 0.12 & 5 \\ 
\hline                                   
\end{tabular}
\newline
\begin{tablenotes}
\item[a]$^\dagger$since explosion epoch t$_0$ =  JD 2457268.5 (2015 September 03 )
     \end{tablenotes}
\label{photometry}
\end{table*}

The spectroscopic observations of SN~2015an commenced with the acquisition of the classification spectrum, followed by 18 epochs of observations (observation log in Table~\ref{tab:spectra_log}) up to 158~d from explosion with the LCO telescopes. The FLOYDS spectrograph at the Faulkes Telescope North and South (FTN and FTS) gives a wavelength coverage of 3200-9000~\AA{} with a resolution ranging from 400-700. 
The 1D wavelength and flux calibrated spectra extraction were carried out using the \texttt{floydsspec} pipeline\footnote{https://www.authorea.com/users/598/articles/6566}. The calibrated spectra were scaled with a factor derived by matching the photometric and spectroscopic continuum flux, thereby correcting for the slit losses. Finally the spectra were corrected for the heliocentric redshift of the host galaxy.

\begin{table}
\caption{Log of the spectroscopic observations.}
\centering
\smallskip
\begin{tabular}{c c c }
\hline \hline
UT Date        & Phase$^\dagger$          & Telescope       \\
                        & (Days)                               &    \\
\hline
2015-09-26.7      & 23.8 & FTS \\
2015-10-01.7      & 28.8 & FTS \\
2015-10-03.7      & 30.8 & FTS \\
2015-10-04.7      & 31.8 & FTS \\
2015-10-06.7      & 33.6 & FTS \\
2015-10-07.7      & 34.8 & FTS \\
2015-10-15.6      & 42.6 & FTN \\
2015-10-25.7      & 52.8 & FTS \\
2015-11-02.5      & 60.6 & FTN \\
2015-11-10.7      & 68.6 & FTS  \\
2015-11-22.7      & 80.8 & FTS \\
2015-11-30.5      & 88.4 & FTN \\
2016-12-08.5      & 96.4 & FTN  \\
2016-12-17.6      & 105.6 & FTS \\
2016-12-25.6      & 113.6 & FTS \\
2016-01-02.4      & 121.4 & FTN \\
2016-01-11.5      & 130.6 & FTN\\
2016-01-31.5      & 150.6 &  FTS \\
2016-02-08.4      & 158.4 & FTS \\
\hline                                   
\end{tabular}
\newline
\begin{tablenotes}
\item[a]$^\dagger$since explosion epoch t$_0$ =  JD 2457268.5 (2015 September 03)
     \end{tablenotes}

\label{tab:spectra_log}      
\end{table}

\section{Light Curve Analysis}
\label{sec4}

\begin{figure}
\includegraphics[scale=0.36, clip, trim={1.8cm 1.8cm 0.65cm 3.9cm}]{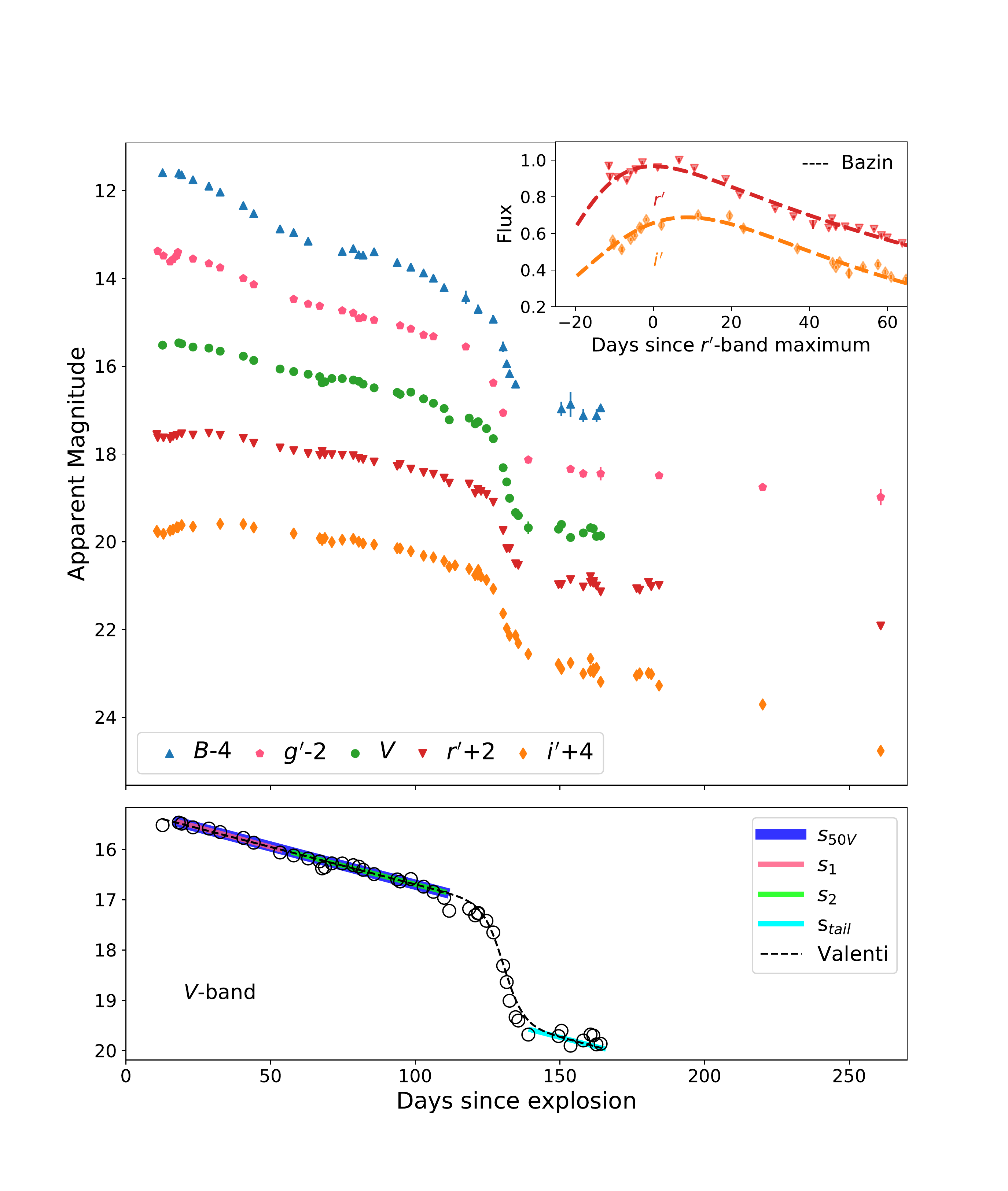}
\caption{$Top~Panel:$ $Bg'Vr'i'$ light curves of SN 2015an offset by a constant value in magnitude in different bands. In the inset plot, the fits to estimate the rise time in the $r'i'$ bands are shown. $Bottom~Panel:$ The fits to the overall decline rate post-maximum up to the fall-off from plateau ($s_{50V}$), early cooling phase ($s_1$), the recombination phase ($s_2$), the radioactive tail phase ($s_{tail}$) and the best-fit to the $V$-band light curve using the function from \citet{2016MNRAS.459.3939V} are shown.}
\label{fig:lc}
\end{figure}

The early light curve (LC) of SN~2015an rises to a peak in the $r^\prime$ and $i^\prime$ bands, while the bluer bands are in the early cooling phase as shown in the top panel of Fig~\ref{fig:lc}. The peak of the redder bands are estimated following the procedure in \cite{2015MNRAS.451.2212G}, where the peak is estimated by fitting a phenomenological function introduced by \cite{Bazin09} to the LC (see eqn. 1 of the paper). The best fit of the functional form is shown in the inset of the top panel of Fig.~\ref{fig:lc}. The rise to maximum from explosion in the $r^\prime$ and $i^\prime$ bands are 16.6 $\pm$ 3.7~d and 25.2 $\pm$ 5.1~d respectively. The multi-band LC changes slope around 55~d since explosion, when the $Vr'i'$ LCs transits to the plateau phase, while $Bg'$ LCs are in a relatively steeper declining phase. The plateau lasts for around 120~d, followed by a sharp fall in brightness after which the SN enters the exponentially declining phase powered by the emission from the decay of $^{56}$Co $\rightarrow$ $^{56}$Fe.

\begin{table}
\caption{The best-fit slopes in the different phases of the $V$-band light curve of SN~2015an measured between start and end phase as well as the best-fit parameters using \citet{2016MNRAS.459.3939V} expression are listed. All slopes are measured in magnitudes per 50d.}
\begin{tabular}{llcc}
\hline\hline
 Slope & Decline rate & Start phase & End phase \\
\hline 
 s$_1$ & 0.78 $\pm$ 0.02 & 15 & 55  \\
 s$_2$ & 0.62 $\pm$ 0.02 & 55 & 120 \\
 s$_{50V}$ & 0.74 $\pm$ 0.01 & 15 & 120 \\
\hline\hline
 $t_{PT}$~(d) & $a_0$~(mag) & $w_0$~(d) & $p_0$~(mag/d)\\
\midrule
130.4 $\pm$ 0.3 &  2.30 $\pm$ 0.05 & 3.4 $\pm$ 0.4 & 0.0148 $\pm$ 0.0004 \\
\hline
\end{tabular}
\label{tab:slopes}   
\end{table}

We performed a linear fit to the $V$-band light curve to estimate the initial steeper decline ($s_1$ = 0.78$\pm$0.02~mag/50~d) and the second shallower decline ($s_2$ = 0.62$\pm$0.02~mag/50~d) before the light curve falls off from the plateau. An overall decline rate, $s_{50V}$ including both these regions (15 to 120~d) is also fitted. We used eqn 1 of \cite{2016MNRAS.459.3939V} to derive other LC parameters such as the drop in magnitude from the plateau to the tail phase ($a_0$), the duration of the transition phase (typically 6 $\times$ $w_0$), the time from explosion to the transition point from the plateau to the tail phase ($t_{PT}$) and the slope of the radioactive tail phase ($p_0$). The best fit parameters are summarized in Table \ref{tab:slopes} and the best fits are shown in the bottom panel of Fig.~\ref{fig:lc}. 

\cite{2014ApJ...786...67A} found the average decline rates for a SNe II sample to be $s_1$=1.32$\pm$0.75 and $s_2$ = 0.64$\pm$0.47 mag/50~d. The decline rates ($s_1$ \& $s_2$) of SN~2015an suggests that this event is consistent with slowly declining SN~II \citep{2014ApJ...786...67A}, while the \cite{2014MNRAS.445..554F} criterion of classifying events with $s_{50V}$\textgreater 0.5 as IIL, places SN~2015an ($s_{50V}$ = 0.74 mag/50~d) under the linearly declining SNe~II. The magnitude drop ($a_0$) of SN~2015an, 2.30$\pm$0.05~mag, lies in the range of 1-2.6~mag as suggested for SNe~II in \cite{2016MNRAS.459.3939V}. 

\begin{table}
\caption{Parameters of the SNe II sample.}
\label{parameter_SNIIP_sample}

\begin{tabular}{@{}lllll}\hline \hline
                    &Parent  & Distance$^\dagger$ & $A_V^{tot}$ & Ref.\\
Supernova& Galaxy & (Mpc)                             & (mag) & \\
\hline

1987A & LMC & 0.05 & 0.60 & 1\\

2005cs & M51 & 7.1 (1.2) & 0.34 & 2,3\\

2009bw & UGC~2890 & 20.2 (0.6) & 0.96 & 4\\

2009N & NGC 4487 & 19.8 (1.1) & 0.403 & 5\\ 

LSQ13fn & LEDA 727284 & 264 (5.3) & 0.167 & 6\\ 

2013by & ESO~138-G10 & 14.74 (1.0) & 0.60 & 7\\

2013ej & NGC~628 & 9.6 (0.7) & 0.19 & 8\\

2017eaw & NGC 6946 & 5.6 (0.1) & 1.271 & 9\\
\hline
\end{tabular}

\flushleft
$^\dagger$ In the H$_0$ = 73.48 km s$^{-1}$ Mpc$^{-1}$ scale.\\
References: (1) \cite{1990AJ.....99.1146H},
(2)~\cite{2006MNRAS.370.1752P},
(3) \cite{2009MNRAS.394.2266P},
(4) \cite{2012MNRAS.422.1122I},
(5) \cite{2014MNRAS.438..368T},
(6) \cite{2016A&A...588A...1P},
(7) \cite{2015MNRAS.448.2608V},
(8) \cite{2015ApJ...807...59H,2015ApJ...806..160B,2016MNRAS.461.2003Y,2016ApJ...822....6D},
(9) \cite{2019ApJ...876...19S}
\end{table}

\begin{figure}
\includegraphics[scale=0.5, width=0.5\textwidth, clip, trim={0.4cm 0.6cm 1.2cm 2.1cm}]{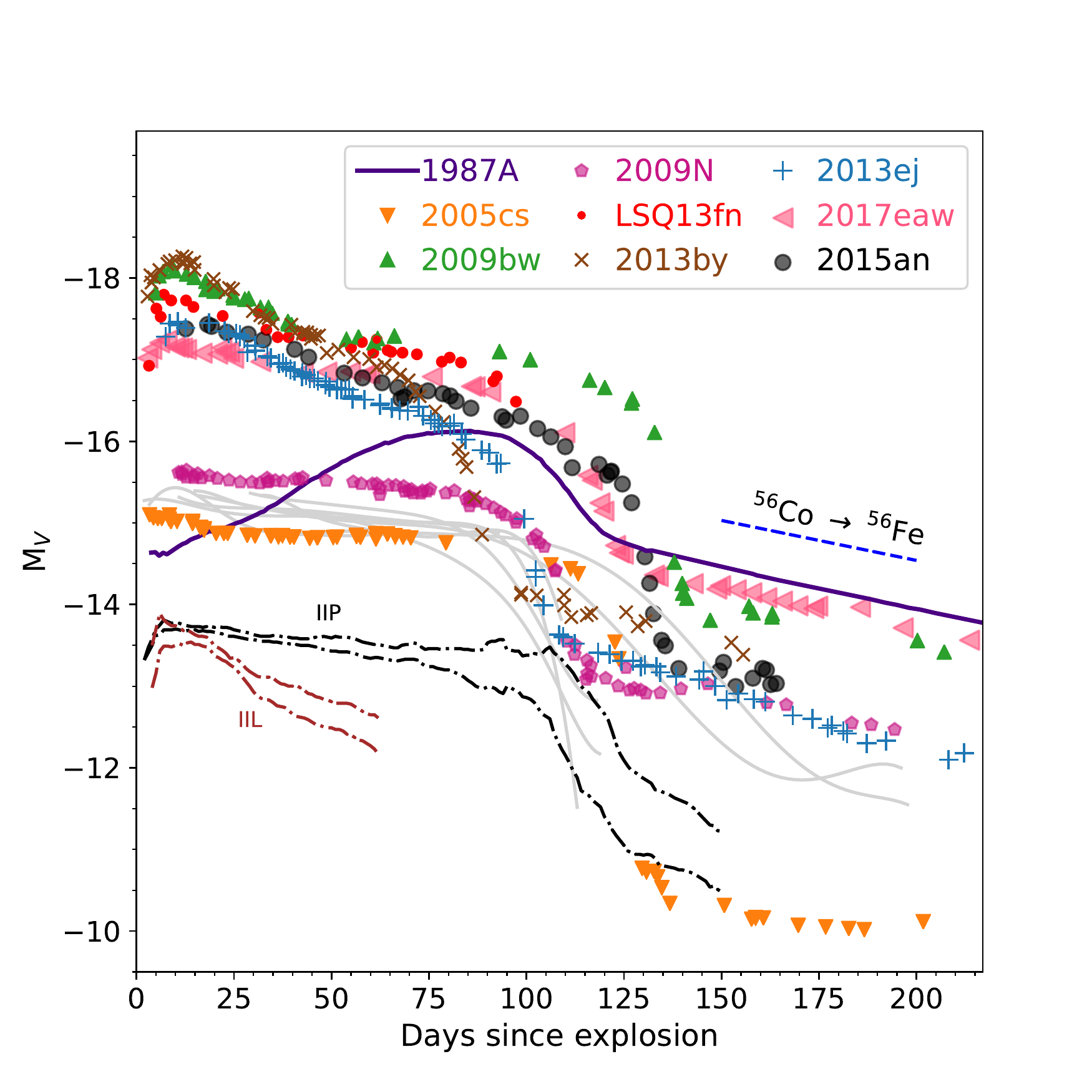}
\caption{Comparison of absolute $V$-band light curves of SN~2015an with other Type II SNe. The magnitudes are corrected for distance and reddening as listed in Table \ref{parameter_SNIIP_sample}. The radioactive decay line assuming full trapping of photons is shown with a dashed line. The dash-dotted lines depicts the range of slopes of Type IIP and IIL SNe as presented in \citet{2014MNRAS.445..554F}. The grey lines are the low-order legendre fits to the light curve of low-luminosity SNe (M$_V$ \textgreater{} $-$15.5~mag) from \citet{2014ApJ...786...67A}}
\label{abs_lc}
\end{figure}

In order to discern the position of SN~2015an in the SNe~II diversity, a sample of events is assembled, to be used as reference and listed in Table \ref{parameter_SNIIP_sample}. The reference sample is composed of -- SN~1987A: the II-pec SN, SN~2005cs representing the low-luminosity events, SN~2009N: a member of the intermediate luminosity events, SN~2009bw: a luminous and long plateau SN IIP, SN~2013by: a prototypical IIL, SN~2013ej: a fast declining IIP/L, LSQ13fn: a SN IIP exhibiting several peculiar characteristics and SN~2017eaw: a prototypical IIP. We have compared the absolute magnitude of SN~2015an with that of the reference sample and low-luminosity SNe (M$_V$ \textgreater{} $-$15.5~mag) from \cite{2014ApJ...786...67A} in Fig.~\ref{abs_lc}, and we see that SN~2015an stands among the brighter SNe II, with an absolute magnitude at 50~d, M$_V^{50}$ = $-$16.83$\pm$0.04~mag, similar to SN~2017eaw. In Fig.~\ref{abs_lc}, the range of slopes possible for the sub-types IIP and IIL from the study of \cite{2014MNRAS.445..554F} is shown with dash-dotted lines and it is clear that the LC of SN~2015an conforms to the IIL class.

\begin{figure}
\includegraphics[scale=0.5, width=0.45\textwidth, clip, trim={0.4cm 0.4cm 0.45cm 0.35cm}]{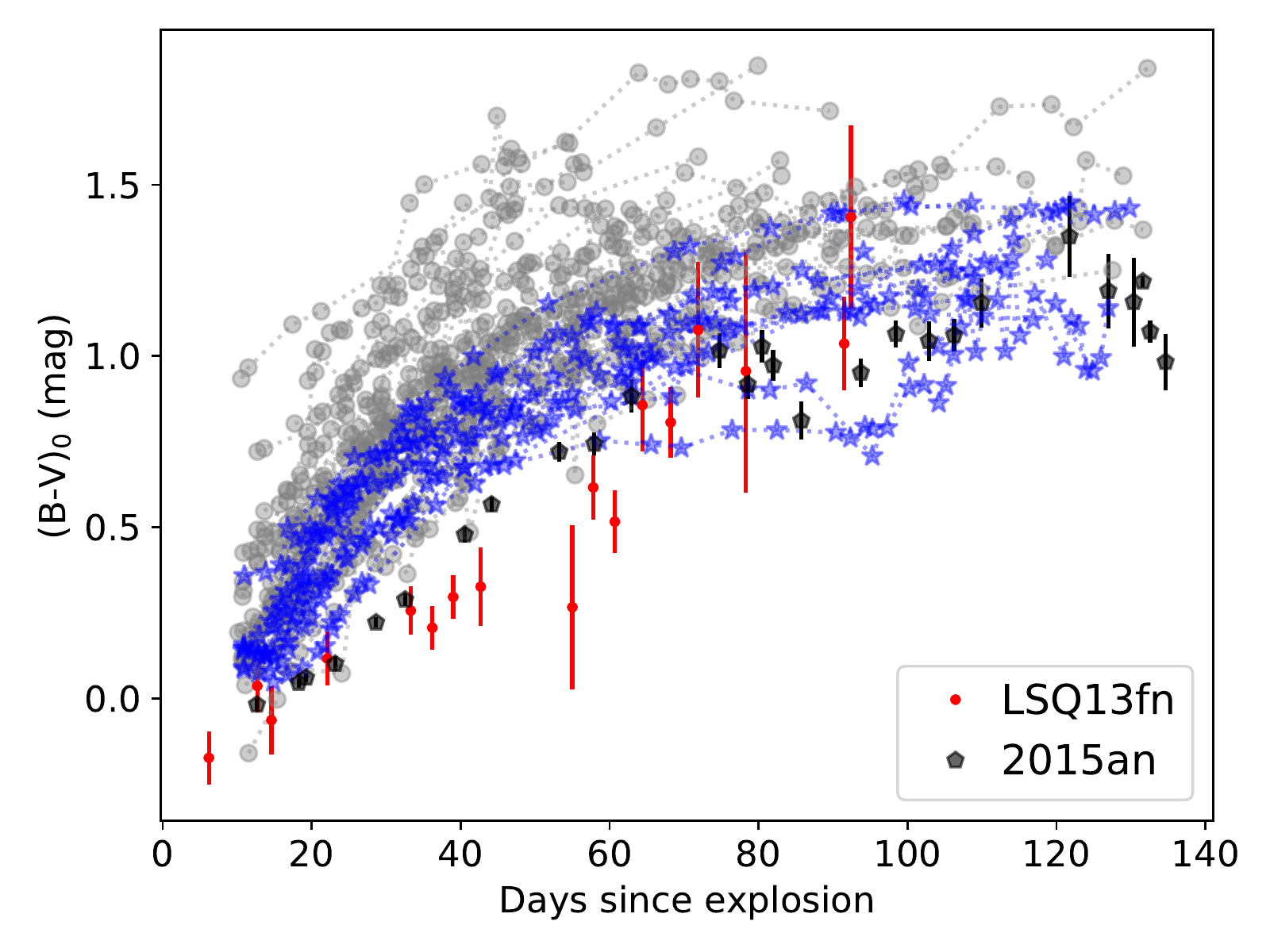}
\caption{The (B-V)$_0$ colour evolution of SN~2015an compared with a sample of SNe II from \citet{2018MNRAS.476.4592D} (shown in grey) and a subset of this sample with negligible host extinction (shown with blue stars). The colour evolution of SN~LSQ13fn is also shown, which falls at the bluer end. All the (B-V) colours are corrected only for Galactic reddening.}
\label{color_lc}
\end{figure}

The ($B-V$)$_0$ colour evolution of SN~2015an is shown in Fig.~\ref{color_lc}, along with the sample of 57 SNe II from \cite{2018MNRAS.476.4592D}. The blue stars are the members of the sub-sample composed of events which are located far away from the host galaxy nuclei and exhibit inconspicuous Na~{\sc i}~D absorption. For a comparison, we also plot the colours of LSQ13fn, which are at the bluer end. Clearly, SN~2015an falls on the bluer edge of the sample, displaying an evolution typical of SNe II. The $(B-V)_0$ colour becomes redder rapidly at early times when the temperature drops quickly following a power law in time, until the commencement of the recombination phase at around 55~d. Since the temperature variation is not much apparent in the plateau phase, colour changes slowly in this phase.

The $^{56}$Ni mass is estimated from the tail bolometric luminosity following \cite{2003ApJ...582..905H} using the tail $V$-band magnitudes. The tail luminosity ($L_t$) obtained from the V-band magnitudes using the bolometric correction factor given in \cite{2003ApJ...582..905H} at 5 epochs -- 158.2, 160.7, 161.7, 162.7 and 164.2~d are 6.11$\pm$0.49~$\times$~10$^{40}$, 6.82$\pm$0.55 $\times$ 10$^{40}$, 6.70$\pm$0.54 $\times$ 10$^{40}$, 5.69$\pm$0.46 $\times$ 10$^{40}$ and 5.75$\pm$0.46 $\times$ 10$^{40}$~erg~s$^{-1}$ respectively, which corresponds to a mean $^{56}$Ni mass of 0.021 $\pm$ 0.010~M$_\odot$. The $^{56}$Ni mass of SNe~2013ej and 2017eaw with similar plateau magnitudes (M$_V^{50}$=$-$16.61 and $-$16.85~mag, respectively) as SN~2015an are 0.018 and 0.045~M$_\odot$, respectively. Thus SN~2015an is more similar to the linearly declining SN~2013ej than the slowly declining SN~2017eaw. The $^{56}$Ni yield of SN~2009N (0.020~M$_\odot$) is the same as SN~2015an, however, the former is an intermediate luminosity SN, with a much lower plateau magnitude ($-$15.59~mag).

\section{Spectral Analysis}
\label{sec5}

\begin{figure}
\includegraphics[scale=0.2,width=0.7\textwidth, clip, trim={8cm 5cm 10cm 9cm}]{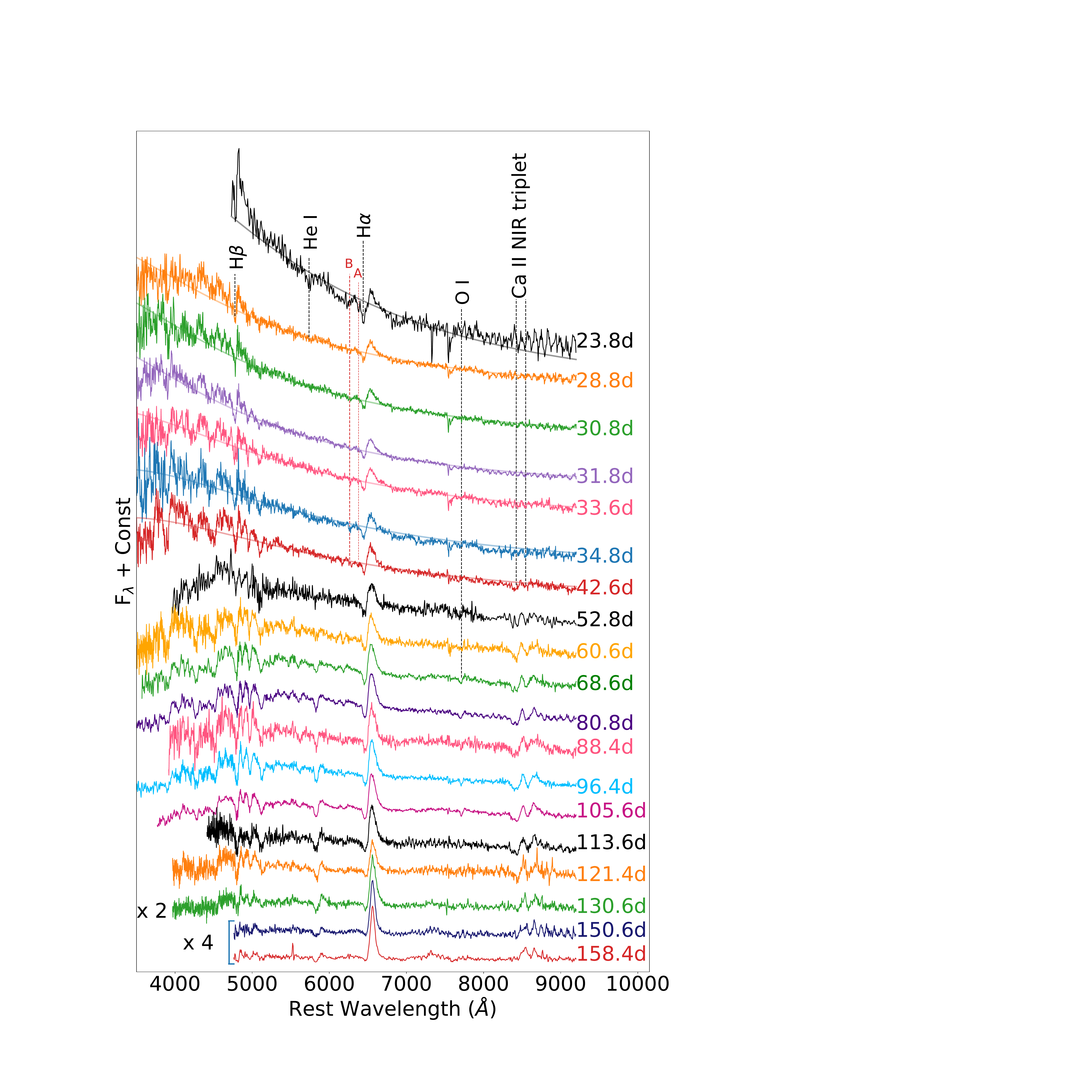}
\caption{The spectral evolution of SN~2015an from 23.8~d to 158.4~d is shown. The spectra are corrected for redshift and Galactic extinction. The conspicuous blue continuum in the early spectra are fitted with a blackbody model to estimate the temperature during these phases. The absorption dips bluewards of the H$\alpha$ profile are marked as \textquoteleft A' and \textquoteleft B'.}
\label{scaled_spectra}
\end{figure}

\begin{figure}
	\begin{center}
		\includegraphics[scale=0.5, width=0.52\textwidth,clip, trim={1.8cm 0.4cm 1.0cm 1.3cm}]{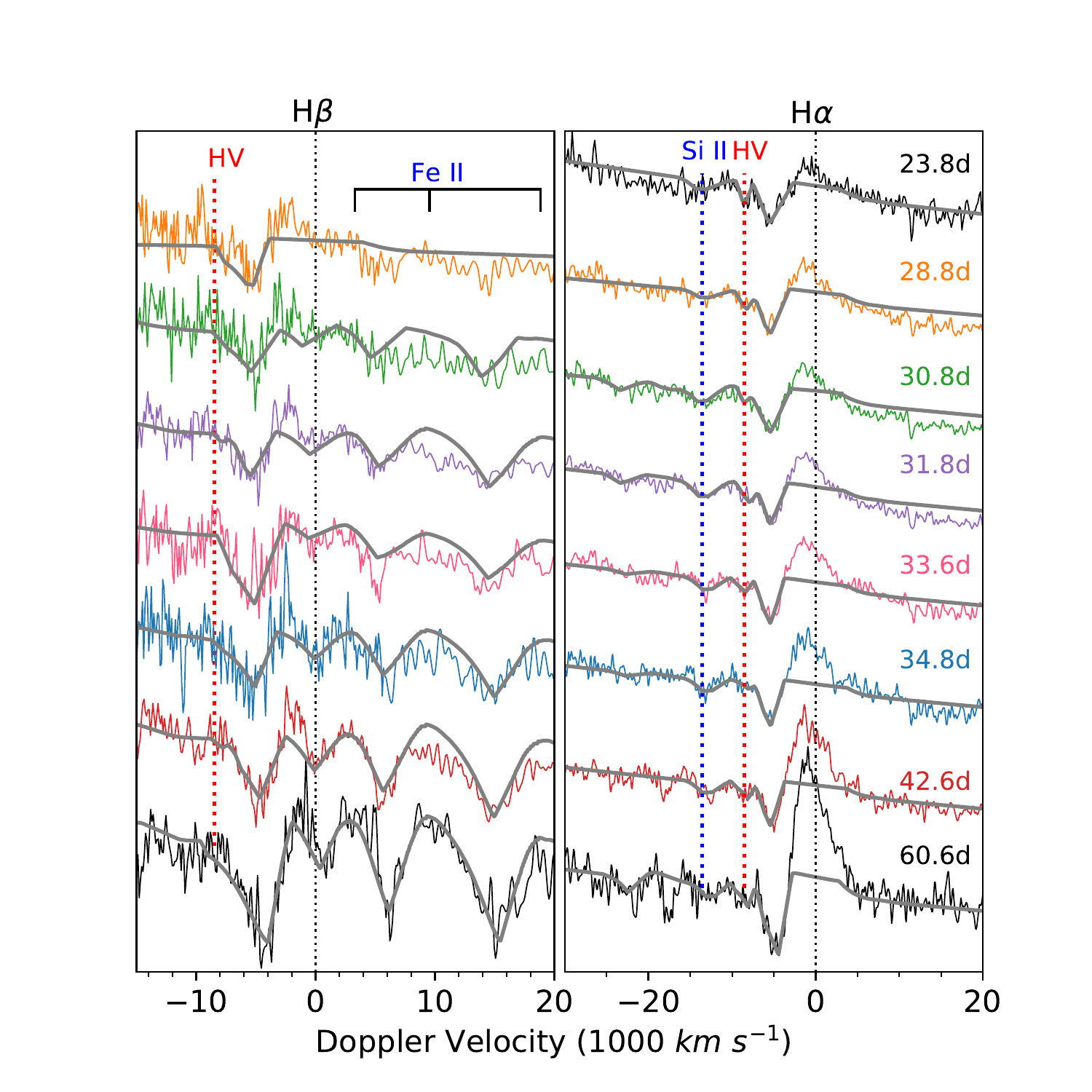}
	\end{center}
	\caption{Evolution of H$\alpha$, H$\beta$ emission line profiles during the photospheric phase. The {\sc syn++} model fits to the spectra are shown with grey solid lines. The Si~{\sc ii} (\textquoteleft B\textquoteright{}) and high velocity (HV) features (\textquoteleft A\textquoteright{}) are conspicuous up to the 50 d spectrum and are shown with dashed lines.}
	\label{fig:vel_space}
\end{figure}

\begin{figure}
\begin{center}
\includegraphics[scale=0.60,clip, trim={0.8cm 0.0cm 1.2cm 1.4cm}]{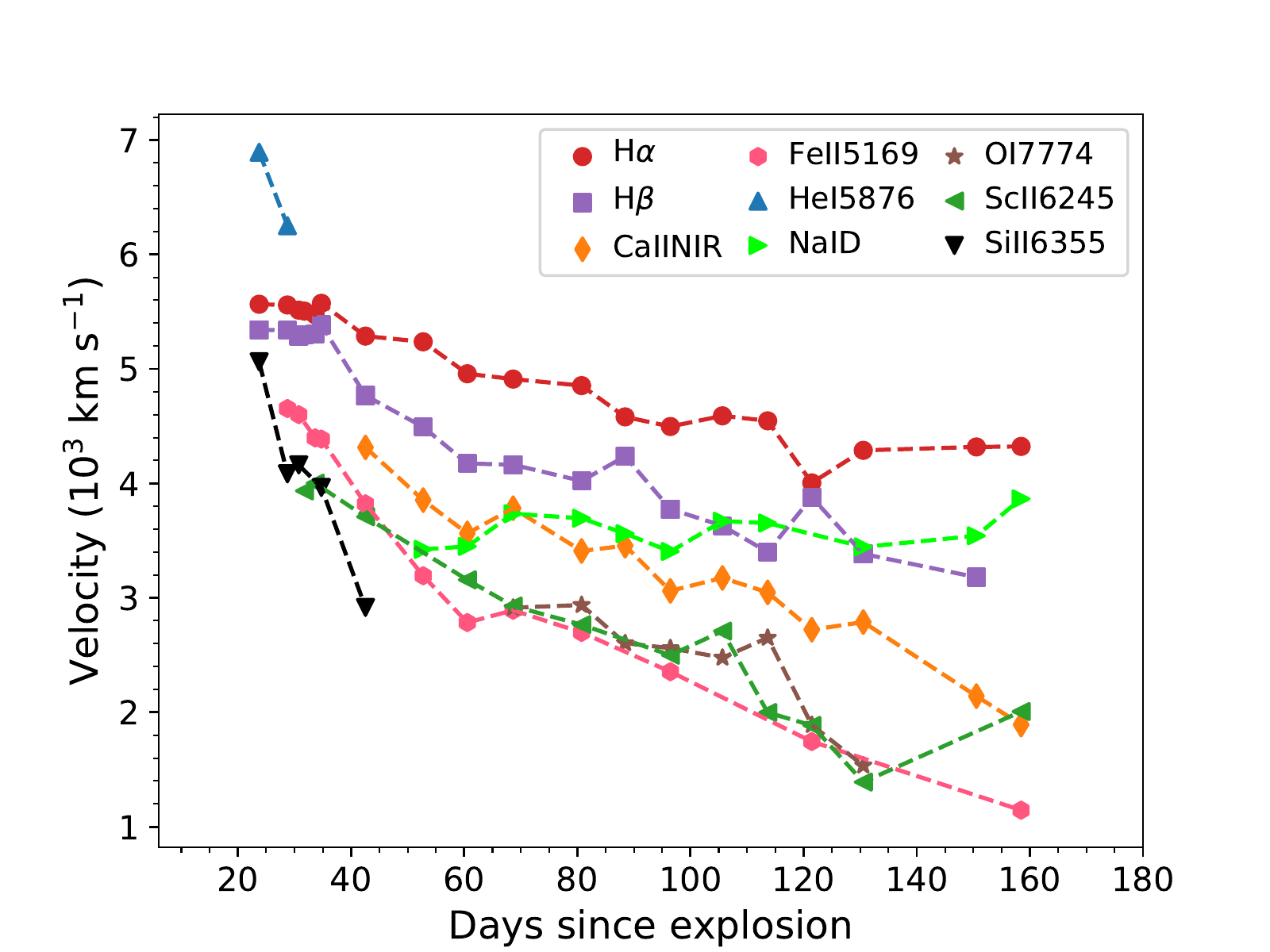}
\end{center}
\caption{Line velocity evolution of H$\alpha$ , H$\beta$ , He {\sc i}, Fe {\sc ii}, Sc {\sc ii}, Ca {\sc ii} NIR, Na {\sc i} D and Si {\sc ii} lines. The velocities are estimated from the absorption minima of these profiles.}
\label{fig:vel_lines}
\end{figure}

Fig.~\ref{scaled_spectra} shows the spectral evolution of SN 2015an from 23.8 to 158.4~d after explosion. The blue continuum in SN~2015an persists longer, up to 42.6~d, than other normal SNe~II. Line identifications were made following \citet{2002PASP..114...35L}. The continuum is fitted with a blackbody function to determine the temperature (shown with solid lines for the first 7 epochs). The temperature is around 8800~K at 28.8~d and then gradually drops to 5600~K at 68.6~d and remains nearly constant afterwards. Prominent H$\alpha$ and weak He~{\sc i}~$\lambda$5876 feature can be discerned in the earliest spectrum. While H$\alpha$ grows with time, the He {\sc i} feature disappears after the 28.8~d spectrum. The Fe~{\sc ii}~$\lambda\lambda\lambda$4094, 5018, 5169 and Ca {\sc ii} H\&K emerge in the 28.8~d spectrum, weak features of Sc~{\sc ii}~$\lambda$6245 and Ca~{\sc ii}~NIR triplet becomes detectable from the 42.6 d spectrum, the Na~{\sc i~d} and Ba~{\sc ii}~$\lambda$6142 features becomes conspicuous from the 60.6 d spectrum and the O {\sc i} $\lambda$7774 lines becomes visible from the 68.6~d spectrum. Fe~{\sc ii} and Sc~{\sc ii} remains visible up to 130.6~d, however the Ba~{\sc ii} and O~{\sc i} lines becomes undetectable after the 105.6~d spectrum. Most of these metal lines disappears in the last two spectra (150.6 and 158.4~d), except Na~{\sc i~d} and Ca~{\sc ii}~NIR triplet.

\begin{figure}
		\includegraphics[scale=1.00, width=0.5\textwidth,clip, trim={0.0cm 0.0cm 0.0cm 0.0cm}]{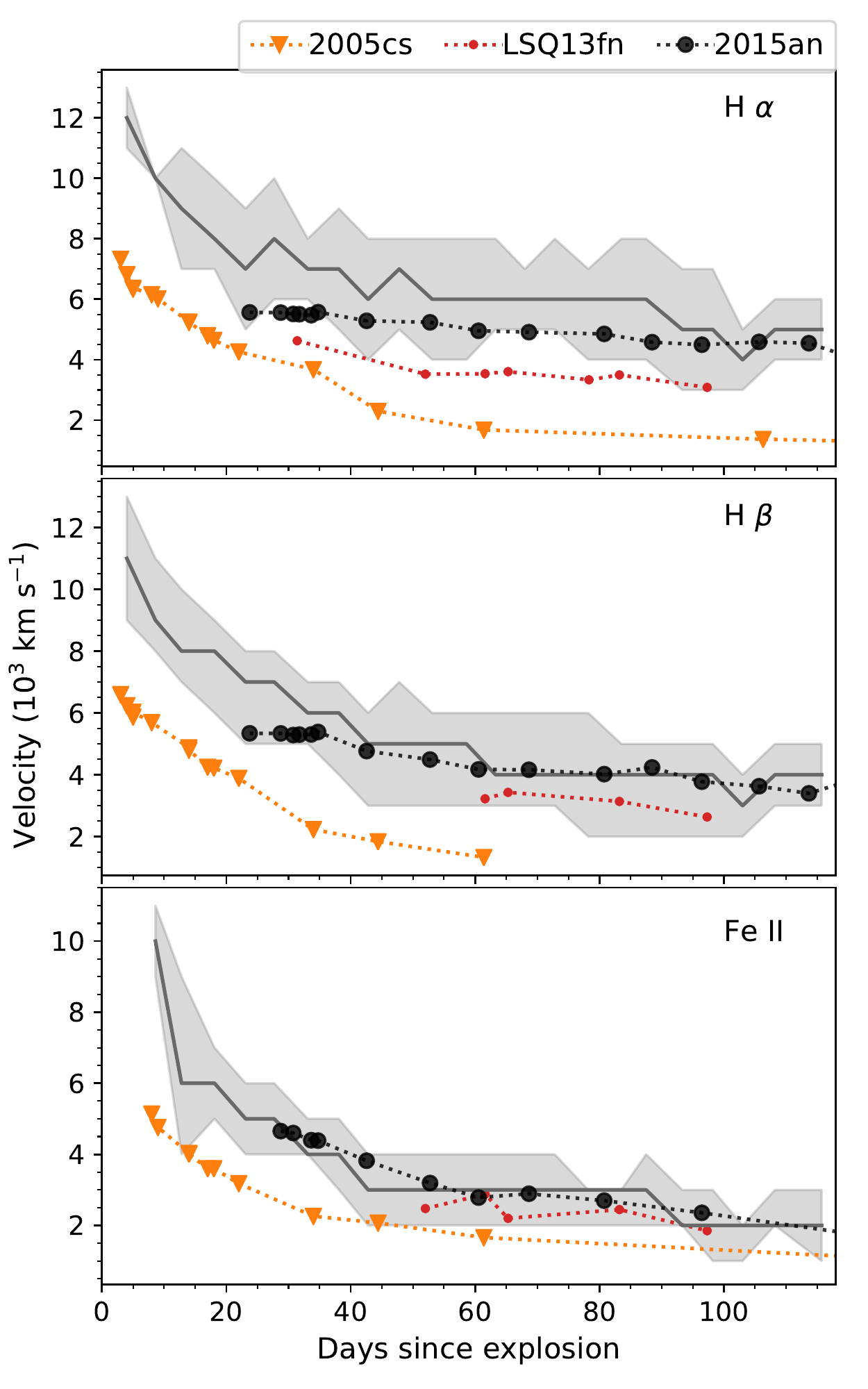}
	\caption{A comparison of the velocity of SN~2015an to SNe~2005cs, LSQ13fn and the mean velocity (grey) of 122 IIP/IIL SNe \citep{2017ApJ...850...90G} for H$\alpha$ (top panel), H$\beta$ (middle panel) and Fe II ($\lambda$5169; bottom panel). The light grey regions represent the standard deviations of the mean velocities of the sample.}
	\label{fig:vel_comp}
\end{figure}

The evolution of H$\alpha$ and H$\beta$ lines from 23.8 to 60.6~d is shown in
Fig.~\ref{fig:vel_space} in the velocity domain and centered at the rest wavelengths of these lines along with the {\sc syn++} \citep{2011PASP..123..237T} fits. The emission peak of H$\alpha$ is shifted by $\sim$ 1100~km~s$^{-1}$ in these phases. Two absortion dips on the blue side of the H$\alpha$ absorption component is visible from the 23.8~d spectrum until 42.6~d and hardly detectable in the 60.6~d spectrum. Such features have been reported at both early and late photospheric epoch in a number of SNe II, and has been termed \textquoteleft cachito\textquoteright{} \citep{2017ApJ...850...89G}. In case of a non-evolving feature, the dip is interpreted to be high velocity (HV) component of H$\alpha$, which usually appears in the late photospheric phases and an evolving component is thought to be originating from Si~{\sc ii}~$\lambda$6355, which typically appears in the early phases. However, rarely these two features have been observed at coeval epochs. For SN~2015an, the feature \textquoteleft A\textquoteright{} does not evolve with time, while the feature \textquoteleft B\textquoteright{} shows an evolving velocity similar to the metal lines (see Fig. \ref{fig:vel_lines}), and hence we identify the feature \textquoteleft A\textquoteright{} as the HV component of H$\alpha$ and the feature  \textquoteleft B\textquoteright{} as Si {\sc ii} $\lambda$6355. This is further confirmed by the {\sc syn++} models fitted to the 23.8 to 60.6~d spectra, where we used a constant HV component ($\sim$8500~km~s$^{-1}$) in addition to the evolving normal velocity component, to reproduce the H$\alpha$ absorption component. While the HV feature is apparent in H$\alpha$, it is inconspicuous in H$\beta$, which is possibly due to the low opacity of H$\beta$ in the circumstellar wind \citep{2007ApJ...662.1136C}, producing rather a broadened absorption feature of H$\beta$.

The velocity evolution of the discernible ions as measured from the shift in the absorption minima from the rest wavelength of these lines are shown in Fig.~\ref{fig:vel_lines}, along with the feature identified as Si {\sc ii} $\lambda$6355. The He {\sc i} line detected in the two earliest spectra has the highest expansion velocity ($\sim$ 6890 and 6250~km~s$^{-1}$ at 23.8 and 28.8~d, respectively) among the ions. The H$\alpha$ and H$\beta$ velocities are nearly similar up to 40~d, and higher than the velocities of the metal lines, suggesting its formation in the outer higher velocity layer of the ejecta. The remarkable characteristic of SN~2015an is, however, the low expansion velocity of H~{\sc i} (5250$\pm$260~km~s$^{-1}$ at 50~d) as compared to SNe~II family (median H$\alpha$ velocity at 50~d post-explosion: 7300~km~s$^{-1}$, \citealt{2017ApJ...850...89G}) with a nearly flat evolution.

\subsection{Comparison with other SNe II}
\subsubsection{Velocity and temperature evolution}

\begin{figure}
\includegraphics[scale=0.5, clip, trim={0cm 0cm 0cm 0cm}]{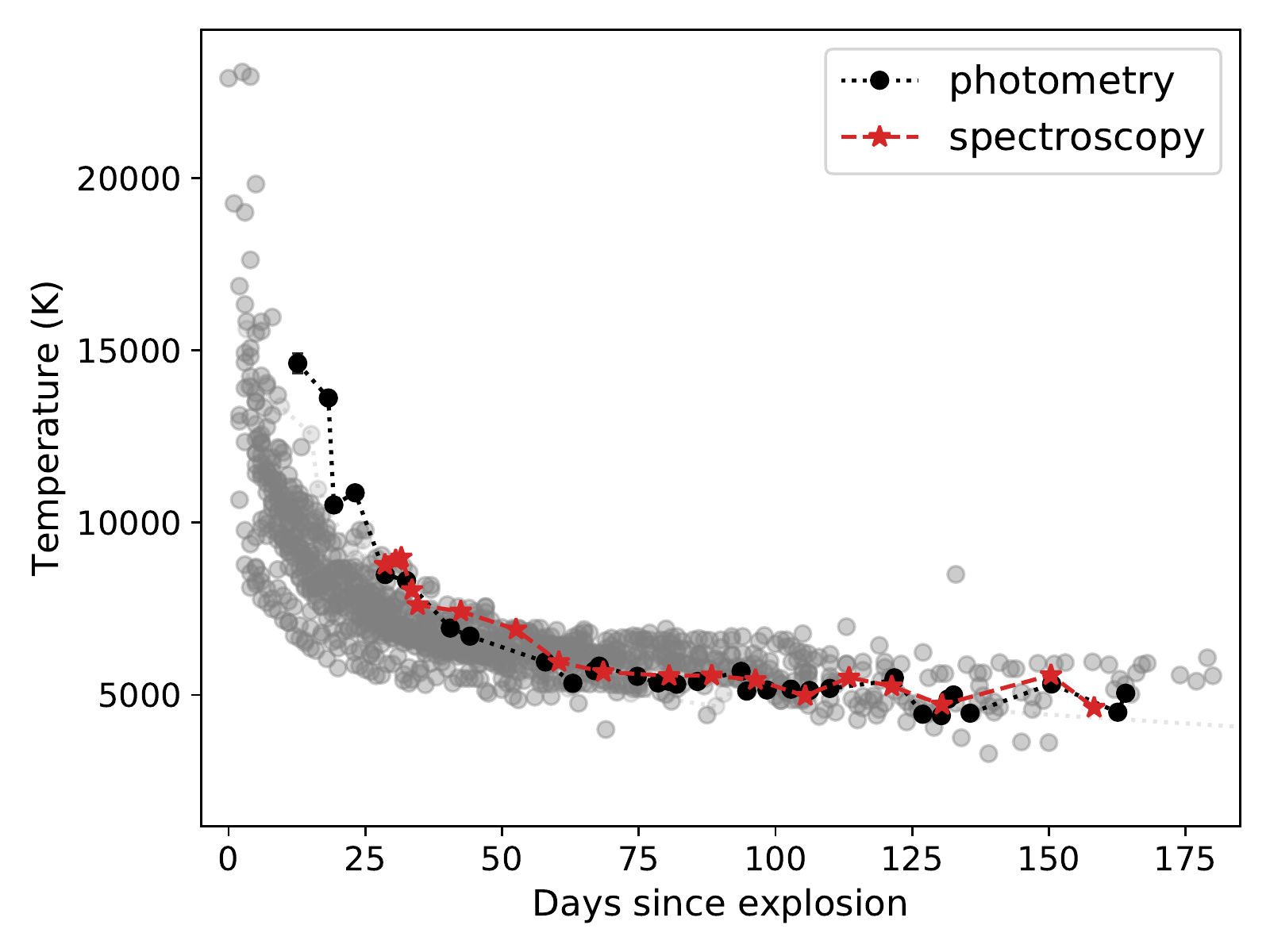}
\caption{Comparison of the temperature evolution of SN~2015an estimated by fitting a blackbody to the photometric and spectroscopic fluxes with a collection of SNe II from \citet{2018MNRAS.473..513F}.}
\label{temp_comp}
\end{figure}

\begin{figure*}
\centering
\includegraphics[scale=1.0, width=1.0\textwidth,clip, trim={3.2cm 0.3cm 3.5cm 2.1cm}]{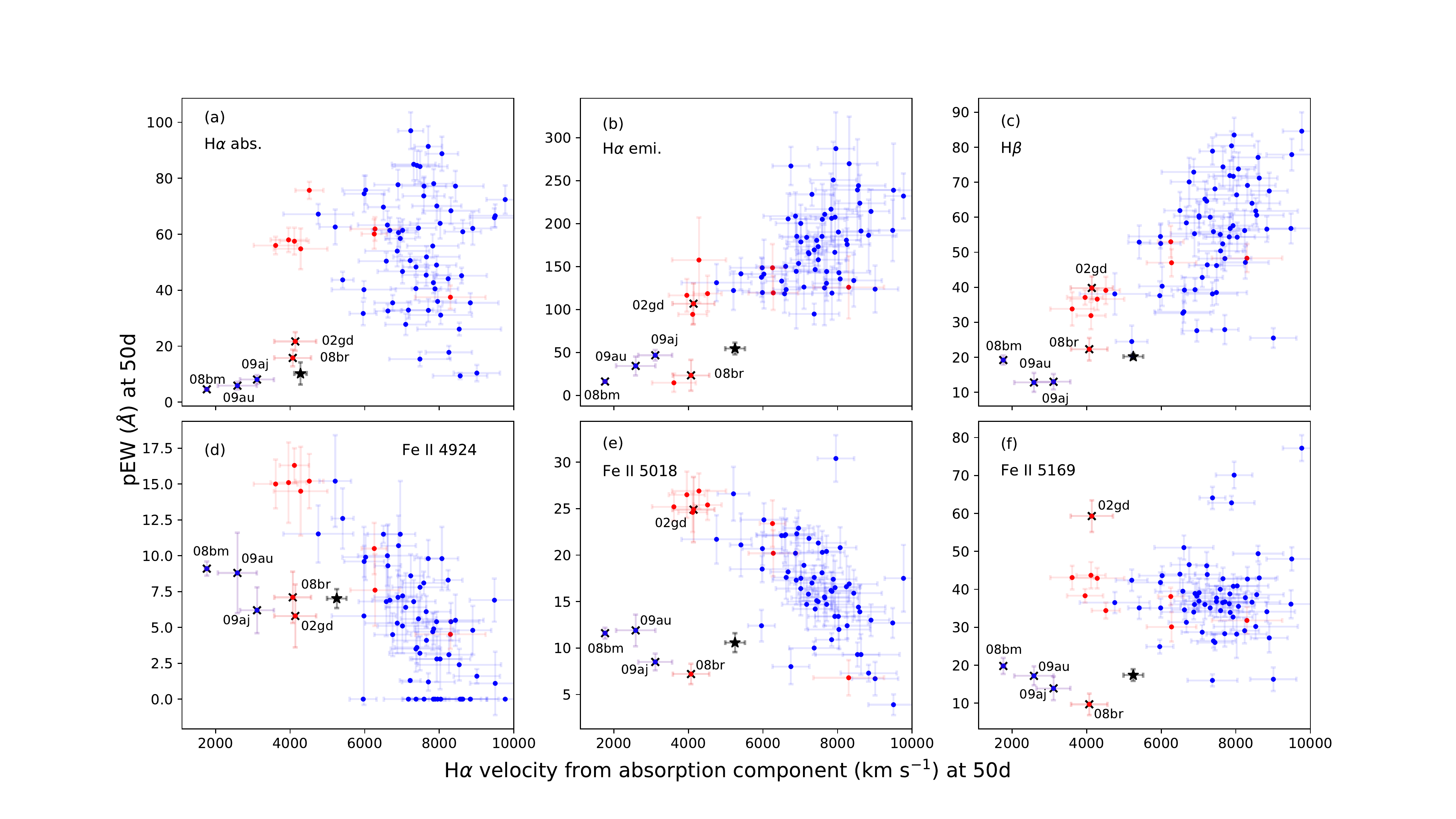}
\caption{The pEW of the prominent features in the 50~d spectra of SN~2015an are plotted against the H$\alpha$ velocity estimated from the shift in the absorption minima at 50~d (marked with $\bigstar$) and compared with the sample of \citet{2017ApJ...850...90G}. The low-luminosity SNe ($-14$ \textgreater{} M$_V^{max}$ \textgreater $-$15.5) are shown in red. SNe~2002gd, 2008bm, 2008br, 2009aj and 2009au, with small pEW of H$\alpha$ and low H$\alpha$ expansion velocity, similar to SN~2015an are marked with \textquoteleft$\times$\textquoteright.}
\label{fig:pEW_vel}
\end{figure*}

Fig.~\ref{fig:vel_comp} shows the H$\alpha$, H$\beta$ and Fe {\sc ii} 5169 velocity evolution of SNe~2005cs, LSQ13fn and 2015an overplotted on the estimated mean velocities of the 122 IIP/L SNe from \cite{2017ApJ...850...90G}. The H$\alpha$ velocity of SN~2015an lies in the lower edge of the velocity range of SNe~II, with significantly low velocity ($\sim$5500~km~s$^{-1}$) at early phases as compared to the median velocity of the sample ($\sim$7900~km~s$^{-1}$). The H$\alpha$ velocity of LSQ13fn falls outside the range of velocities of SNe II in the sample, but is still higher than the low-luminosity SN~2005cs by $\sim$800~km~s$^{-1}$ and $\sim$1700~km~s$^{-1}$ at early (\textless 30~d) and late phases (\textgreater 40~d), respectively. The H~{\sc i} line velocities of SN~2015an does not vary much and shows a nearly flat evolution. Nevertheless, the H$\beta$ and Fe~{\sc ii} velocity evolution of SN~2015an matches with the mean velocity evolution of the sample, except in the early phase (\textless 30~d), where the velocity of H$\beta$ falls on the lower limit of the mean velocity.

The temperature evolution of SN~2015an, obtained by performing blackbody fit to the SED constructed from photometric and spectroscopic fluxes, is shown in Fig.~\ref{temp_comp}. The grey points depicts the evolution of temperature of a collection of 29 SNe II from \cite{2018MNRAS.473..513F}. Initially the temperature of SN~2015an (14600~K at 12.8~d) is higher than other SNe II at similar phases. The temperature of the sample at $\sim$13~d varies in the range of 6580 to 10840~K with a median temperature of $\sim$9000~K. Later than 40~d, when the SN enters the recombination phase, the temperature of SN~2015an is $\sim$7400~K still around 1000~K higher than the median temperature ($\sim$6500~K) of the sample. The blue (B-V) colour and the excess blue flux in the early spectra corroborates with the high temperature of SN~2015an in the early phase. 

\subsubsection{Comparison of plateau spectra}

\begin{figure*}
\begin{center}
\includegraphics[scale=0.8,clip, trim={3.5cm 1.5cm 13cm 3cm}]{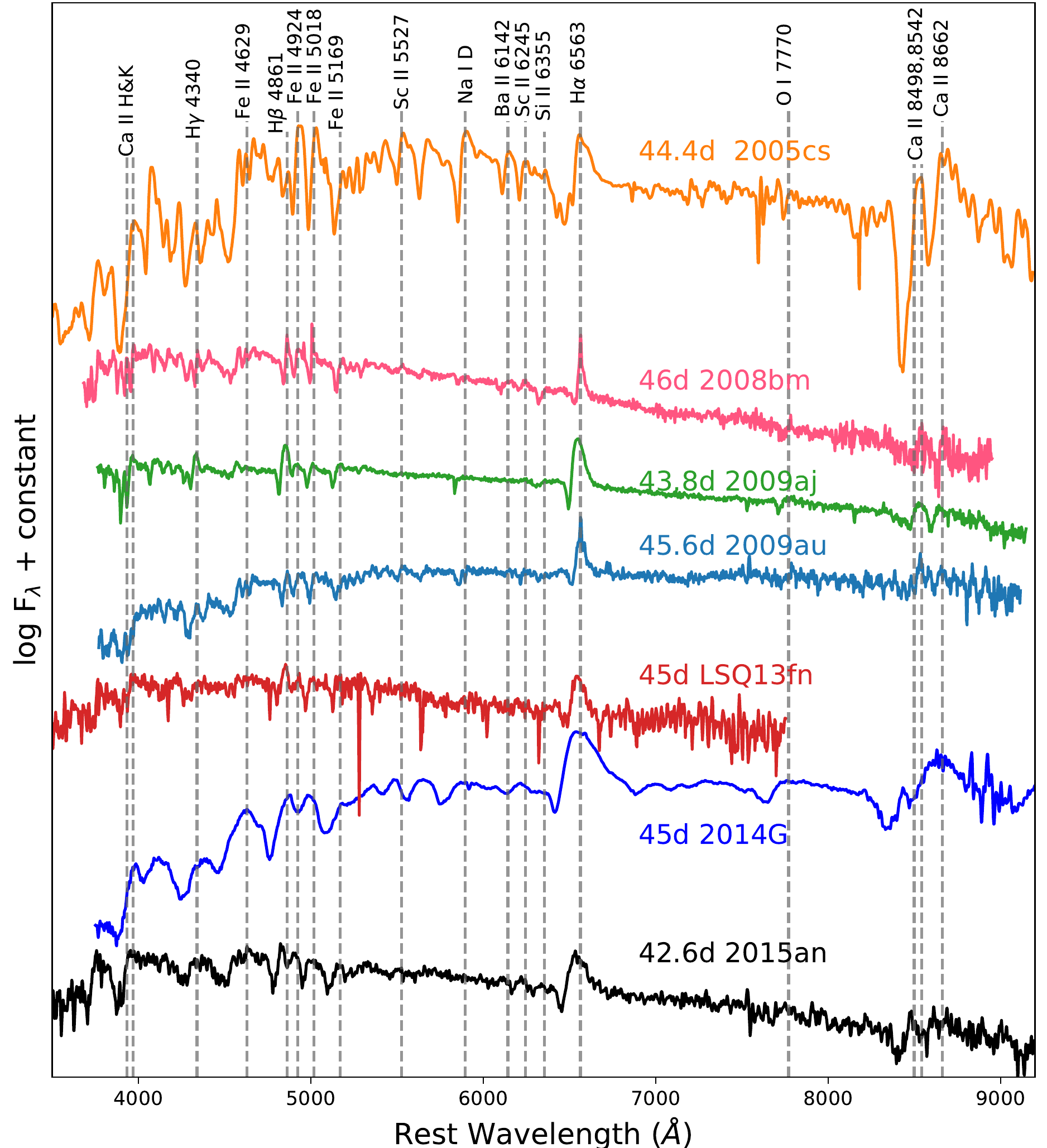}
\end{center}
\caption{Comparison of mid-plateau spectrum (42.6 d) of SN 2015an with other SNe from the reference sample.}
\label{fig:spec_comp}
\end{figure*}

\begin{figure}
\begin{center}
\includegraphics[scale=0.4,clip, trim={1.5cm 2.0cm 0cm 3.5cm}]{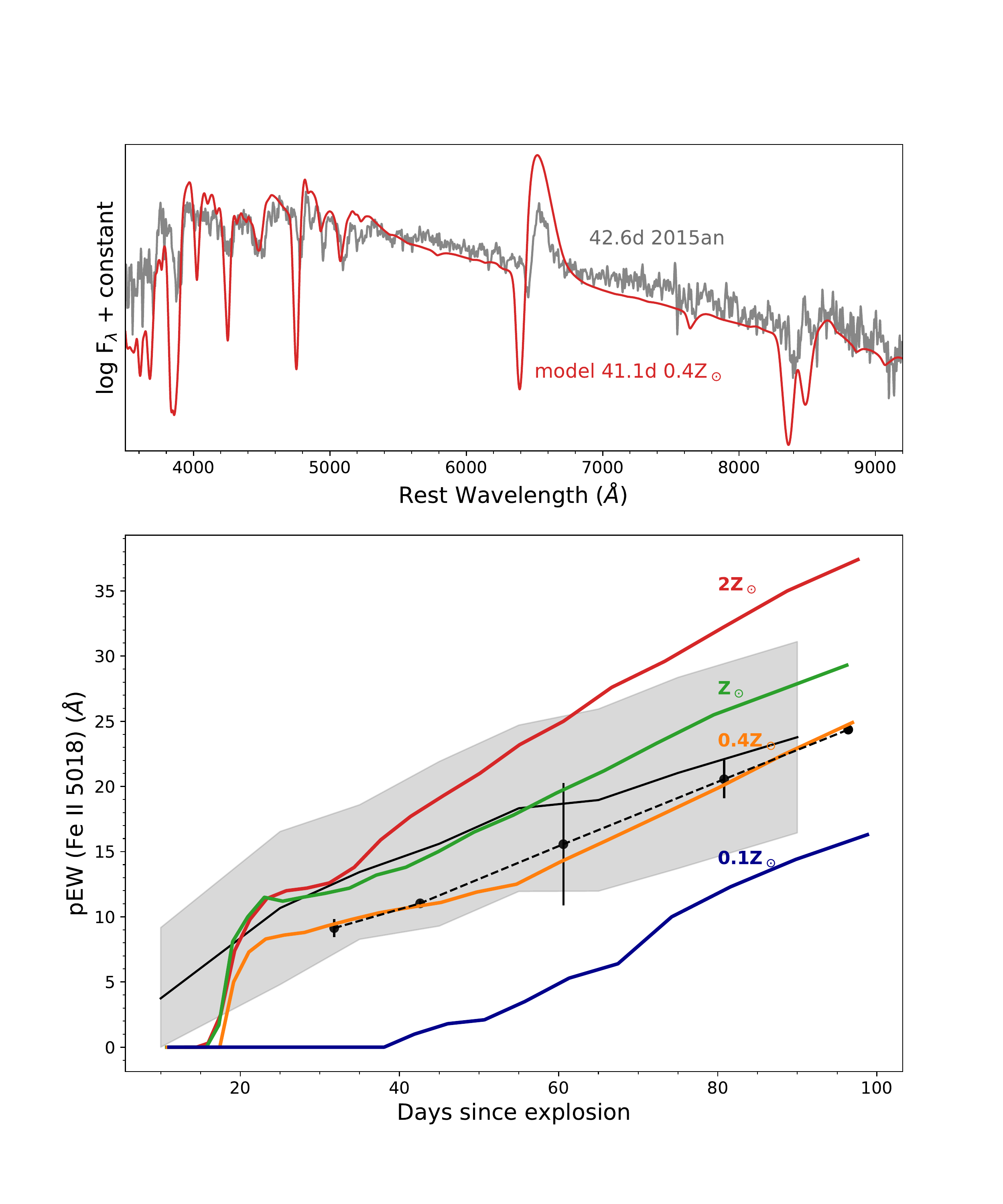}
\end{center}
\caption{{\it Top Panel:} Comparison of a mid-plateau spectrum (42.6 d) of SN~2015an with the 0.4~Z$_\odot$ model spectrum of \citet{2013MNRAS.433.1745D}. {\it Bottom Panel:} The evolution of the pEW of Fe~{\sc ii}~$\lambda$5018 of SN~2015an is shown with dashed lines. The black line is the mean pEW of a sample of 119 SNe~II from \citet{2016A&A...589A.110A} and the shaded region shows the standard deviation about the mean. The time evolution of the pEWs of Fe~{\sc ii}~$\lambda$5018 of the four distinct metallicity models of \citet{2013MNRAS.433.1745D} is shown with thick lines.}
\label{fig:spec_comp_mod}
\end{figure}

SN~2015an displays some peculiar characteristics, in addition to the usual properties of SNe~II, thereby necessitating the cognizance of the status of SN~2015an in the SNe~II population before heading for the comparison with a sample of events. To this aim, we plotted the pseudo-equivalent widths (pEW) of the prominent lines vs H$\alpha$ velocity at 50~d of SN~2015an along with a sample of 122 SNe~II from \cite{2017ApJ...850...90G} in Fig.~\ref{fig:pEW_vel}. This SN~II sample exhibits a range of H$\alpha$ expansion velocities (1500~\textless~v$_{H\alpha}$~\textless~9600~km~s$^{-1}$) and pEWs of lines, presumably linked to the diversity in the explosion energies, progenitor radius and metallicity of SNe~II. As noted in \cite{2017ApJ...850...90G}, the H$\alpha$ velocity exhibits the strongest positive correlation with the pEW of the emission component of H$\alpha$ and the absorption component of H$\beta$ and negative correlation with pEW of Fe~{\sc ii}~$\lambda$4924 and Fe~{\sc ii}~$\lambda$5018. In panel (a), SN~2015an lies at the lower left corner (small pEW and low v$_{H\alpha}$), along with SNe~2002gd, 2008br, 2008bm, 2009aj and 2009au, clearly far off from the majority of SNe~II. However, in panels (b) and (c), SN~2015an seems to fall in the trend of a positive correlation of expansion velocity with pEW of H$\alpha$ emission component and H$\beta$ absorption component, respectively. In the case of Fe~{\sc ii} $\lambda\lambda\lambda$4924,5018,5169 (panels (d), (e) and (f)), SN~2015an lies intermediate to that of majority of SNe II and the four SNe on the left. Out of the five SNe from the lower left corner in panel (a), SNe 2008bm, 2009au and 2009aj shows unusually low velocities for their brightness and narrow emission lines at early phases, indicative of CSI at play. So, for comparison, we use the $\sim$45~d dereddened and de-redshifted spectra of SNe 2008bm, 2009aj and 2009au along with that of the low-luminosity SN~2005cs, the Type IIP SN~LSQ13fn and the Type IIL SN~2014G. The spectra of SN~2015an obtained on 42.6~d and the comparison sample are shown in Fig.~\ref{fig:spec_comp}. While SN~2005cs shows well-developed metal lines, all other SNe exhibit comparatively weaker metal lines. The continuum of SN~2015an matches well with that of SN~LSQ13fn and neither of them has a conspicuous Na~{\sc i}~D line at this phase, as compared to the other SNe. The SNe displaying weaker metal lines has enhanced continuum flux bluewards of 5200~\AA, compared to SNe~2005cs and 2014G except SN~2009au. 

\cite{2014MNRAS.440.1856D} suggested that the SN progenitor metallicity plays a crucial role in the evolution of metal lines in the photospheric phase in that the SNe with low progenitor metallicity exhibit weaker metal features. \cite{2013MNRAS.433.1745D} simulated model spectra by evolving a 15~M$_\odot$ ZAMS star up to the pre-SN stage at different metallicities (0.1, 0.4, 1.0, 2.0~Z$_\odot$). Comparing the 42.6~d spectrum of SN~2015an with these models we found the best match with 0.4 Z$_\odot$ model spectrum (pEW of Fe~{\sc ii}~$\lambda$5018 in the 42.6~d spectrum of SN~2015an is 11.0~\AA{} and that of the model spectrum at 41.1~d is 10.6~\AA), which is overplotted on the spectrum of SN~2015an in the top panel of Fig.~\ref{fig:spec_comp_mod}. Moreover, using a sample of 119 host H~{\sc ii} regions  \cite{2016A&A...589A.110A} estimated the oxygen abundances and found a positive correlation between the host H~{\sc ii} region oxygen abundance and pEW of Fe~{\sc ii}~$\lambda$5018. In the bottom panel of Fig.~\ref{fig:spec_comp_mod}, the time evolution of the mean pEW of Fe~{\sc ii}~$\lambda$5018 of the sample of SNe~II from \cite{2016A&A...589A.110A} is shown along with its standard deviation (the black line and the shaded region), over which the thick solid lines are the time evolution of the pEW of Fe~{\sc ii}~$\lambda$5018 in the different metallicity models of \cite{2013MNRAS.433.1745D}. The pEW measurements of Fe~{\sc ii}~$\lambda$5018 in the spectra of SN~2015an is shown with a dashed line which traces the evolution of the  0.4~Z$_\odot$ model and is lower than the mean pEW of the sample.

\section{Distance}
\label{dist}
\begin{figure}
\includegraphics[scale=0.55, width=0.44\textwidth,clip, trim={0.4cm 0.4cm 0.4cm 0.3cm}]{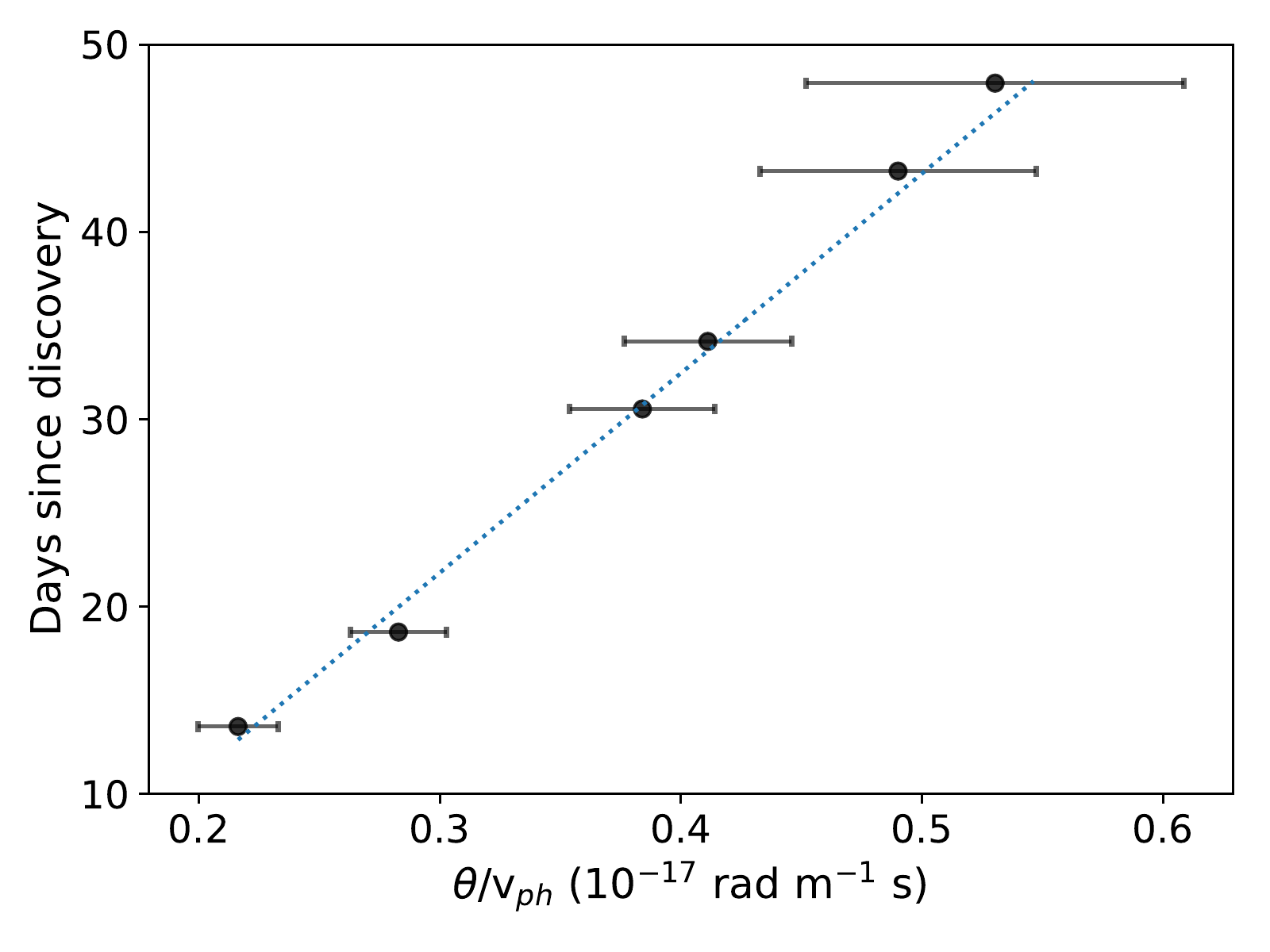}
\caption{Distance determination with the EPM for SN 2015an, using \citet{2005A&A...439..671D} prescription for the dilution factors and the $BV$ observations. The distance and the explosion epoch are obtained from the best fit.}
\label{fig:EPM}
\end{figure}

We implement the EPM to estimate the distance from early photometric and spectroscopic observations as outlined in \cite{2018MNRAS.479.2421D}. The EPM rests on the assumption that the ejecta is homologously expanding and radiating as a diluted blackbody. Thus, the colour temperature (T$_c$) and the angular radius ($\theta$) were obtained by fitting diluted blackbody function to the spectral energy distribution constructed from the observed magnitudes, where the dilution factors were adopted from \citet{2005A&A...439..671D}. Moreover, the filter response function needs to be convolved with the model blackbody function, to eliminate its effect from the observed broadband magnitudes. The convolved function can be expressed as a function of T$_c$ and the coefficients were adopted from \cite{2001ApJ...558..615H}. The photospheric velocities (v$_{ph}$) were estimated from the minima of the absorption profile of H$\alpha$ in the first epoch (13.6~d from discovery) and Fe~{\sc ii}~$\lambda$5169 in the rest of the epochs (up to $\sim$ 50~d from discovery).

The distance and the time of explosion $t_0$ can then be estimated from the slope and y-intercept, respectively, of the following expression:
\begin{equation}
t =D(\theta/v_{ph})+t_0
\end{equation}
The linear fit to the data is shown in Fig.~\ref{fig:EPM} and the distance and explosion epoch estimates from the fit are 29.8~$\pm$~1.5~Mpc and 2457268.5 $\pm$ 1.6 d (2015~September~03~UT), respectively. 

\section{Semi-analytical Modelling of the bolometric light curve}
\label{sec7}
To generate the observed bolometric light curve, first the observed fluxes were corrected for distance and reddening using the values given in Table~\ref{tab:sn15an_IC2367_detail}. Then SED at different epochs was constructed encompassing the fluxes from the UV to the IR wavelengths. Due to the unavailability of UV and IR data, we extrapolated the SED constructed in the optical bands, approximating it as a blackbody, to the UV and IR bands, following the same prescription as used in the direct integration method in \cite{2017PASP..129d4202L}. At late phases, a linear extrapolation is performed in the UV regime. The decline rate of the radioactive tail in the bolometric light curve is found to be 0.0140~mag/d which is steeper than 0.0098~mag/d, indicating incomplete trapping of the $\gamma$-rays produced by the radioactive decay of $^{56}$Co $\rightarrow$ $^{56}$Fe.

We used semi-analytical modelling of \cite{2014A&A...571A..77N,2016A&A...589A..53N} to model the bolometric light curve of SN~2015an. To synthesize the fast initial decline and the late phase of SNe II, this model invokes a two component structure of ejecta, where a massive core is surrounded by a low mass envelope. 
\begin{figure}
\begin{center}
\includegraphics[scale=1.0, width=0.53\textwidth,clip, trim={0.7cm 13.3cm 0.2cm 2.9cm}]{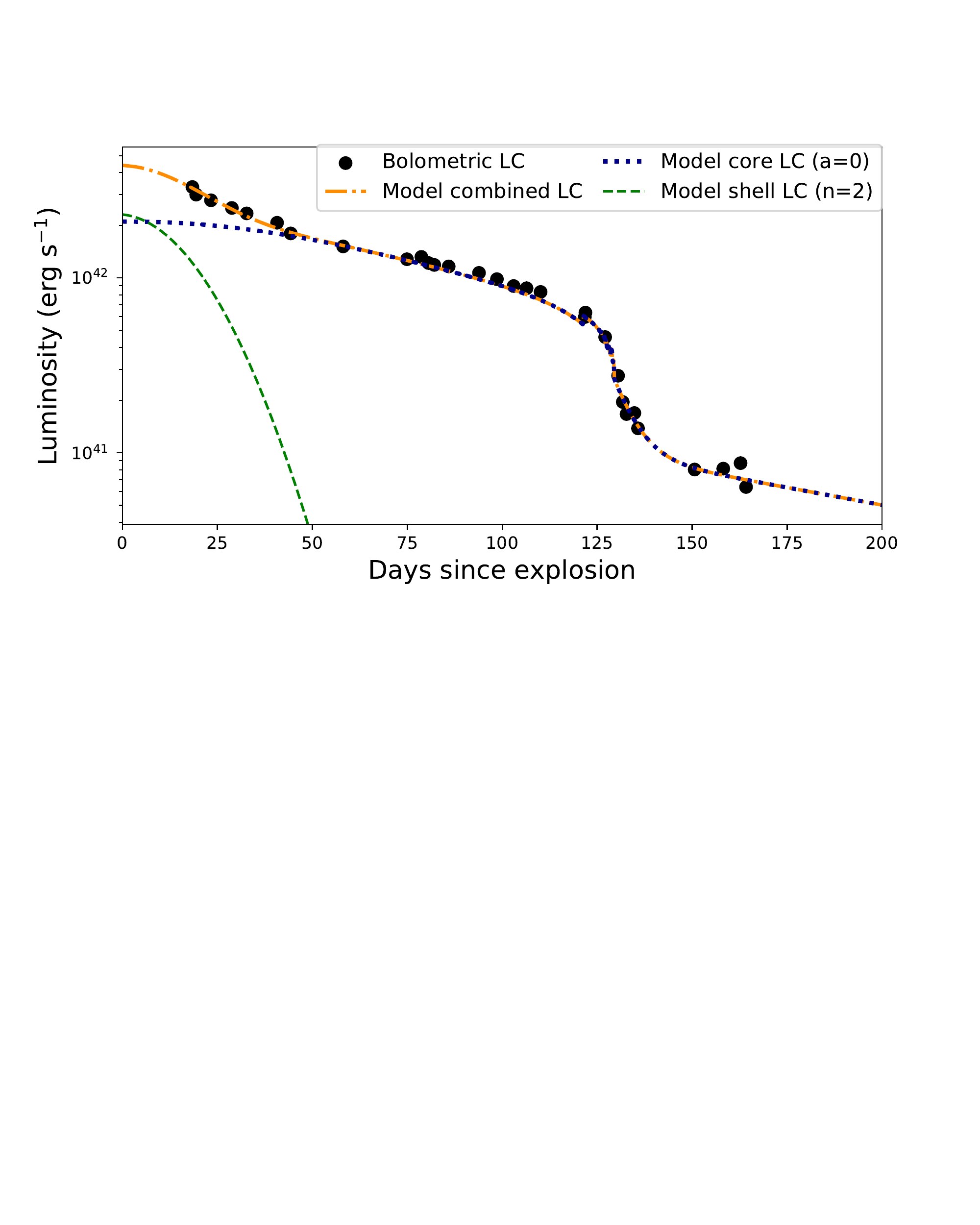}
\end{center}
\caption{The observed bolometric LC of SN~2015an and the best fit two component analytical model of \citet{2016A&A...589A..53N}.}
\label{nagy}
\end{figure}

\begin{figure}
\begin{center}
\includegraphics[scale=1.0, width=0.5\textwidth,clip, trim={0.4cm 0.55cm 0.2cm .2cm}]{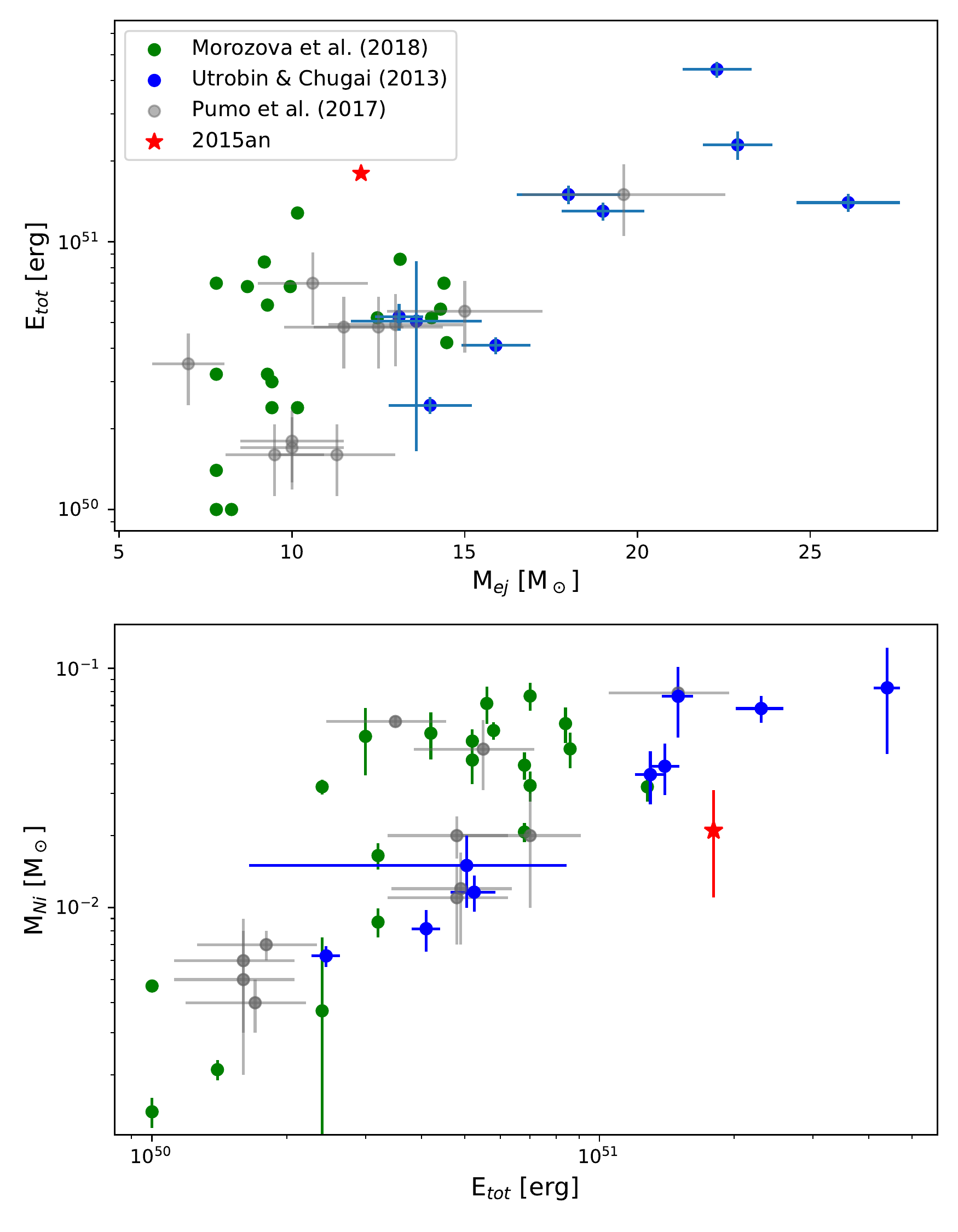}
\end{center}
\caption{The total explosion energies of SNe (E$_{tot}$ = E$_{kin}$+E$_{th}$) are plotted against the ejecta masses and the ejected $^{56}$Ni masses of a sample of SNe from \citet{2013A&A...555A.145U,2017MNRAS.464.3013P,2018ApJ...858...15M}}
\label{Efin_Mej}
\end{figure}

We used a constant density profile for the core and a power law density profile for the envelope. The opacity values for the core and the shell are kept fixed to the average opacity recommended for SNe IIP ($\kappa_{core}$=0.2~cm$^2$ g$^{-1}$ and $\kappa_{shell}$=0.38~cm$^2$ g$^{-1}$, \citealt{2018ApJ...862..143N}). The estimated best fit parameters are the initial radius (2.7$\times$10$^{13}~$cm$\sim$388~R$_\odot$), ejecta mass M$_{ej}$ (12~M$_\odot$) and total energy (1.8~foe). Assuming the remnant mass to be 1.5-2.0~M$_\odot$, the minimum ZAMS mass of the progenitor is 14~M$_\odot$. The best-fit model is over-plotted on the bolometric LC in Fig.~\ref{nagy} and the parameters for the core and the shell are listed in Table~\ref{Nagy}. In the top panel of Fig.~\ref{Efin_Mej}, the total energy of SN~2015an (E$_{tot}$ = E$_{kin}$+E$_{th}$) is plotted against the ejecta mass, along with that derived for other SNe using hydrodynamic modelling \citep{2013A&A...555A.145U,2017MNRAS.464.3013P} and SNEC models \citep{2018ApJ...858...15M}. SN~2015an falls in the trend of the positive correlation between the two parameters. In the bottom panel of Fig.~\ref{Efin_Mej}, the $^{56}$Ni mass is plotted against E$_{tot}$, where we find that compared to its energy, the $^{56}$Ni mass yield should have been higher. 

\begin{table}
\setlength{\tabcolsep}{2pt}
 \begin{minipage}{84mm}
  \caption{The best fit core and shell parameters for true bolometric light curve of SN 2015an using \citet{2016A&A...589A..53N}.}
  
  \begin{tabular}{@{}lccc@{}}
  \hline
  \hline
Parameter & Core & Shell & Remarks  \\
 \hline
R$_0$ (cm)  & 2.7 $\times$ 10$^{13}$ & 9.0 $\times$ 10$^{13}$ & Initial radius of ejecta\\
T$_{rec} (K)$ & 7000 & -  & Recombination temperature\\
M$_{ej}$ (M$_\odot$) & 12 & 0.3 & Ejecta mass\\
E$_{th} (foe)$ & 0.9 & 0.04 & Initial thermal energy \\
E$_{kin} (foe)$ & 0.9 & 0.06 & Initial kinetic energy \\
M$_{Ni}$ (M$_\odot$) & 0.02 & - & Initial $^{56}$Ni mass\\
$\kappa$ (cm$^2$ g$^{-1}$) & 0.2 & 0.38 &Opacity\\
A$_g$ (day$^2$) & 3.0 $\times$ 10$^{5}$ & 1 $\times$ 10$^{10}$ & Gamma-ray leakage \\
\hline
     \label{Nagy}
     \end{tabular}
\end{minipage}
\end{table}

\section{Discussion}
\label{sec8}
\begin{figure*}
\begin{center}
\includegraphics[scale=1.0, width=1.0\textwidth,clip, trim={0.4cm 0.4cm 0.2cm 0.3cm}]{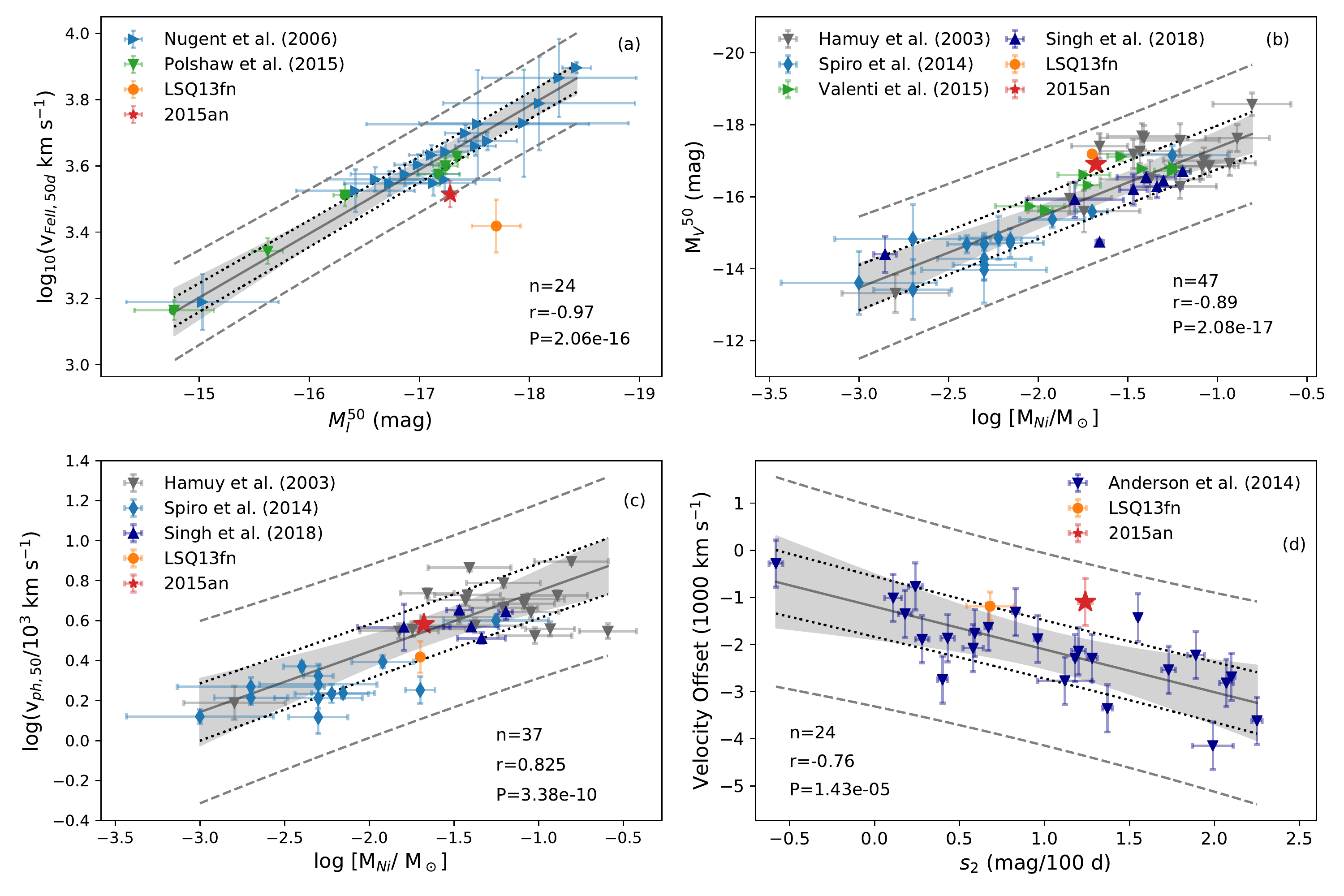}
\end{center}
\caption{The collocation of SN~2015an in the correlation plots among absolute magnitude at 50~d (M$_I^{50}$), the photospheric velocity at 50 d (v$_{ph,50}$), $^{56}$Ni mass, H$\alpha$ velocity offset from emission peak at 30~d and the decline rate of the V-band light curve in plateau phase (s$_2$) in mag/100d is shown along with the sample of SNe~II from \citet{2003ApJ...582..905H,2006ApJ...645..841N,2014MNRAS.439.2873S, 2014MNRAS.441..671A,2015A&A...580L..15P,2016MNRAS.459.3939V, 2018MNRAS.480.2475S}. The shaded region shows the 3$\sigma$ confidence interval, while the dotted and dashed line shows the 1 and 3$\sigma$ prediction limits, respectively. For each panel, n is the number of events, r is the Pearson correlation coefficient, and P is the probability of detecting a correlation by chance.}
\label{cor_plots}
\end{figure*}
As discussed above, SN~2015an exhibits some peculiar features in addition to the regular properties of SNe~II. The H$\alpha$ expansion velocity of SN~2015an in the early phases is comparatively lower than other SNe~II of similar luminosity. High velocity component of H$\alpha$ is conspicuous at the early phases, which is not commonly observed in early SN~II spectrum. Moreover, the persistence of the excess in flux bluewards of 5200~\AA{} up to $\sim$ 43~d, is also atypical of SNe~II. We discuss below these properties of SN~2015an and its possible implications.

\subsection{Collocation of SN~2015an in the SN~II parameter space}
In Fig.~\ref{cor_plots}, we use samples of SNe II from recent studies to discern the position of SN~2015an in the SN~II diversity. The regression line for the literature sample, excluding SNe~LSQ13fn and 2015an, is shown in grey, the shaded region gives the 3$\sigma$ confidence interval for the regression coefficients and the dotted and dashed lines are the 1$\sigma$ and 3$\sigma$ prediction interval, respectively, for the regression model. The Pearson correlation parameters listed in the figure are also estimated using the literature samples. 

In panel (a) of Fig.~\ref{cor_plots}, the expansion velocity of Fe~{\sc ii}~$\lambda$5169 and absolute magnitude in the $I$-band of SN~2015an (obtained by converting the $i$-band magnitude to $I$-band magnitude using the transformation equation in \cite{2006A&A...460..339J}) at 50~d are shown, along with that of SN~LSQ13fn and using the sample of \cite{2006ApJ...645..841N} and \cite{2015A&A...580L..15P}. SN~LSQ13fn lies outside the 3$\sigma$ prediction band, which  implies that this event has a much lower expansion velocity as compared to its luminosity. SN~2015an, on the other hand, is fainter than SN~LSQ13fn by $\sim$0.4~mag and with an expansion velocity higher by 600~km~s$^{-1}$ than SN~LSQ13fn, manages to settle itself right over the 3$\sigma$ prediction limit. 

In Fig. (b), the absolute magnitude in $V$-band at 50~d is plotted against the logarithm of $^{56}$Ni mass for a sample of SNe from the literature \citep{2003ApJ...582..905H, 2014MNRAS.439.2873S, 2016MNRAS.459.3939V, 2018MNRAS.480.2475S}. Both SNe~LSQ13fn and 2015an lies just outside the 1$\sigma$ prediction limit but well within the 3$\sigma$ prediction interval.

In Fig. (c), the photospheric velocity is plotted against the logarithm of $^{56}$Ni mass for a sample of SNe \citep{2003ApJ...582..905H, 2014MNRAS.439.2873S, 2018MNRAS.480.2475S}. In this case, we find both SNe~LSQ13fn and 2015an lie within the 1$\sigma$ prediction interval. 

In Fig. (d), the emission peak velocity offset of H$\alpha$ at 30~d is plotted against the decline rate in the plateau phase (s$_2$) using the sample from \cite{2014MNRAS.441..671A}, where the faster declining SNe have higher offsets. This is expected as faster declining SNe has a steeper ejecta density structure, enhancing the occultation of the receding part of the ejecta, thereby shifting the emission peak to the bluer wavelengths. With a decline rate ($s_2$) of 1.24$\pm$0.04~mag/100~d, the emission peak of H$\alpha$ in  SN~2015an has a velocity offset smaller by about 1000~km~s$^{-1}$ than the mean offset of the sample, possibly due to its low expansion velocity. 

Thus, the normal luminosity of SN~2015an in conjunction with its lower expansion velocity and lower $^{56}$Ni mass yield, suggests the existence of an external source boosting the luminosity of this event, most likely the interaction between the ejecta and CSM. 

\subsection{Early Circumstellar Interaction}
For SN~2015an, we do not have any spectrum in the first three weeks after explosion. The first spectrum obtained on $\sim$23.8~d is dominated by blue continuum. Moreover, a notch bluewards of the H$\alpha$ absorption feature is clearly discernible up to 42.6~d spectrum (Fig.~\ref{fig:vel_space}). This feature is identified as the high velocity feature (HV) of H$\alpha$ at 8500~km~s$^{-1}$ considering its non-evolving nature and further substantiated in the {\sc syn++} modelling. The HV feature is attributed to the excitation of the outer recombined part of the ejecta with X-rays from the reverse shock as a result of the interaction of ejecta and circumstellar material \citep{2007ApJ...662.1136C}. SN~2015an also shows bluer colour and higher temperature, in comparison to a sample of SNe II at coeval epochs (Figs.~\ref{color_lc}, \ref{temp_comp}), which also indicates an interaction of SN ejecta with a nearby CSM, converting the ejecta kinetic energy to thermal energy. 

The HV feature, however, is not apparent in the early phase of most SNe II. The appearance of this feature in SN~2015an at early times may be attributed to the low expansion velocity of hydrogen. A low velocity ejecta will take longer time to cover the nearby CSM. Once the CSM is traversed by the ejecta, the HV feature is expected to diminish. It may again reappear as the ejecta becomes transparent, depending on the density of the CSM. 

\subsection{Low metallicity progenitor}
The other remarkable feature of SN~2015an is the presence of weak metal lines in the photopheric phase, particularly Fe~{\sc ii}~$\lambda\lambda$5018,5169. \cite{2013MNRAS.433.1745D} showed that metallicity plays a crucial role in metal line formation in SN spectra, during the recombination phase, when the photosphere essentially samples the outer H envelope, that is relatively less affected by mixing or nuclear burning. While the pEW of the H~{\sc i} line will be dependent on both the CSI and metallicity, as CSI will broaden the absorption component and shallow H~{\sc i} lines are expected to form from a higher metallicity ejecta, the metal lines will be solely affected by the metallicity of the progenitor. Moreover, the excess of flux bluewards of 5200~\AA{} in SN~2015an, in comparison to other SNe II (such as SNe~2005cs and 2014G) at coeval epochs, indicates a less efficient line blanketing in SN~2015an. These properties suggest a low-metallicity progenitor for SN~2015an. The progenitor metallicity can potentially affect the colour of SNe II as well. The sample study of host H~{\sc ii} region of SNe II to obtain metallicity estimates by \cite{2016A&A...589A.110A}, suggests a strong correlation of the pEW of Fe~{\sc ii}~$\lambda$5018 with metallicity. We found SN~2015an to match best with the 0.4~Z$_\odot$ metallicity model of \cite{2013MNRAS.433.1745D}. This indicates that SN~2015an possibly originated from a sub-solar metallicity progenitor. Furthermore, since metallicity gradients in galaxies are found to be radially decreasing outwards \citep{1999PASP..111..919H}, the location of SN~2015an at a deprojected radial distance of 13.9~kpc, suggests a lower metallicity environment for the progenitor of SN~2015an.

Finally, it is worth noting that some atypical features of SN~2015an, such as its higher photospheric temperature and persistence of blue spectrum later than usual, is related to the explosion time. Since we do not have early spectra of SN~2015an (obtained within two weeks of discovery), we could not impose strong pre-explosion constraints. If SN~2015an was discovered soon after explosion (than our adopted explosion epoch corresponding to 10~d before discovery), then the temperature and colour of SN~2015an would become more similar to normal SNe~II. Nevertheless, a spectrum of SN~II with a dominant blue continuum two weeks past discovery and normal luminosity in concurrence with low expansion velocity of H$\alpha$, as in case of SN~2015an, is unusual.

\section{Summary}
\label{sec9}
In this paper, we present the photometric and spectroscopic analysis of SN~2015an in the galaxy IC~2367. The striking feature of SN~2015an is its low H$\alpha$ expansion velocity in the early phases as compared to its brightness. The absolute magnitude of SN~2015an at 50~d is M$_V^{50}$ = $-$16.83$\pm$0.04~mag, which places this event among the brighter SNe~II. The slope of the $V$-band light curve in the first 50~d is 0.78$\pm$0.02 mag, which is steeper than the slowly declining SNe~II.  The colour evolution of SN~2015an follows the same trend as SNe~II family, however, this event is bluer in comparison to the rest of the sample. The Ni mass derived from the tail luminosity is 0.021$\pm$0.010~M$_\odot$. 

The spectra of SN~2015an is apparently similar to the SNe~II population with prominent H~{\sc i} P~Cygni profiles, however the metal lines are comparatively weaker and the blue continuum lasts longer than typical SNe~II. High velocity features have also been identified in the early spectra of SN~2015an, suggesting a possible role of CSI. The temperature derived from the photometric and spectroscopic SEDs is relatively higher than SNe~II in the early phases, implying the conversion of the kinetic energy of ejecta to thermal energy in the ejecta-CSM interaction. The weak metal lines and the blue continuum is possibly associated with the metallicity of the progenitor, as sub-solar metallicity progenitor produces weaker metal lines \citep{2013MNRAS.433.1745D}. 

We used the expanding photosphere method and derived a distance of 29.8$\pm$1.5~Mpc to SN~2015an, which is in accord with the Virgo infall distance. The explosion parameters derived from the best fit semi-analytic model generated using the prescription of \cite{2016A&A...589A..53N} to the bolometric light curve yielded a total ejecta mass of $\sim$12~M$_\odot$, a total explosion energy of 1.8 foe and an initial radius of 388~R$_\odot$. The estimated initial radius is small as compared to other RSGs, and hence SN~2015an was expected to show faster cooling with a rapid transition from the early cooling phase to the plateau phase and a redder continuum. We infer that the combined effect of ejecta-circumstellar interaction and a low-metallicity progenitor is giving rise to the peculiar properties of SN~2015an.
Furthermore, it becomes apparent that CSI is not only important for Type IIn or some IIL, but also low-velocity SNe~II may have some interaction.

\section*{Acknowledgments}
We thank Thomas de Jaeger, J. P. Anderson, C. P. Guti$\grave{e}$rrez and V. P. Utrobin for sharing data. This research has made use of the NASA/IPAC Extragalactic Database (NED) which is operated by the Jet Propulsion Laboratory, California Institute of Technology, under contract with the National Aeronautics and Space Administration. We acknowledge the usage of the HyperLeda data base (http://leda.univ-lyon1.fr). DAH, CM, and GH were supported by NSF-1313484. This work makes use of data obtained by the LCO network. Research by SV is supported by NSF grants AST-1813176. SBP and KM acknowledges BRICS grant DST/IMRCD/BRICS/Pilotcall/ProFCheap/2017(G) for the present work. SBP and KM also acknowledge
the DST/JSPS grant, DST/INT/JSPS/P/281/2018. KM acknowledges the support from Department of Science and Technology (DST), Govt. of India and Indo-US Science and Technology Forum (IUSSTF) for the WISTEMM fellowship and Dept. of Physics, UC Davis where a part of this work was carried out.




\bibliographystyle{mnras}
\bibliography{refag}   



\appendix



\bsp	
\label{lastpage}
\end{document}